\documentclass[final,5p,times,twocolumn]{elsarticle}

\graphicspath{{figs1/}}

\usepackage{amssymb}
\usepackage{amsmath}

\usepackage{graphicx}
\usepackage{subcaption}
\usepackage[labelfont=bf]{caption}
\usepackage{float}
\usepackage{hyperref}
\hypersetup{
    colorlinks=true,
    linkcolor=blue,
    filecolor=magenta,      
    urlcolor=blue,
}
\usepackage{placeins}
\usepackage{bm,csquotes}
\usepackage{enumerate}
\usepackage{booktabs}
\usepackage{algorithm}%
\usepackage{algorithmicx}%
\usepackage{algpseudocode}%

\usepackage{siunitx}
\usepackage{multirow}

\makeatletter
\renewcommand\paragraph{\@startsection{paragraph}{4}{\z@}%
  {1ex \@plus 1ex \@minus .2ex}%
  {-1em}%
  {\normalfont\normalsize\bfseries}}
\makeatother

\journal{Artificial Intelligence In Medicine}

\begin{document}

\begin{frontmatter}



\title{Hybrid Probabilistic Forecasting of Under-Five Malaria Admissions in Ghana: A Gaussian Process Regression with Holt-Winters Smoothing}

\author[label2]{T. Ansah-Narh\corref{cor1}}
\cortext[cor1]{Corresponding author.}
\ead{theophilus.ansah-narh@gaec.gov.gh}

\author[label1]{Y. Asare Afrane} 

\author[label2]{J. Bremang Tandoh} 



\affiliation[label2]{organization={Ghana Space Science and Technology Institute, Ghana Atomic Energy Commission},
            addressline={P. O. Box LG 80}, 
            city={Legon},
            state={Accra},
            country={Ghana}}
            
\affiliation[label1]{organization={Department of Medical Microbiology, University of Ghana Medical School}, 
            city={University of Ghana},
            state={Accra},
            country={Ghana}}




\begin{abstract}
Accurate malaria forecasting remains a major challenge in sub-Saharan Africa, where strong seasonality, reporting uncertainty, and non-stationary transmission dynamics reduce the reliability of conventional epidemiological time-series models. In Ghana, district-level malaria surveillance requires forecasting frameworks that are not only predictive but also probabilistically rigorous, operationally interpretable, and robust under limited-data conditions. This study proposes a hybrid probabilistic forecasting framework that integrates Gaussian Process Regression (GPR) with Holt--Winters exponential smoothing for modelling monthly malaria admissions among children under five years of age. Within this framework, GPR captures complex non-linear temporal behaviour and predictive uncertainty, while Holt--Winters smoothing stabilises long-horizon forecast trajectories and preserves recurrent seasonal structure.
Using ten years of district-level malaria surveillance data (2014--2023), model performance was evaluated using a rolling-origin expanding-window validation framework to ensure assessment on sequential unseen future observations. Full-sample analyses were additionally performed to examine each model’s ability to reproduce the underlying temporal dynamics of malaria transmission. The proposed hybrid model achieved substantially improved predictive performance ($R^2 = 0.9906$) relative to the best standalone baseline model ($R^2 = 0.8213$ for Holt--Winters), while maintaining well-calibrated residual behaviour, with approximately $94.2\%$ of residuals contained within $\pm 2\sigma$ uncertainty bounds. The results further demonstrate that integrating probabilistic machine learning with seasonal smoothing reduces forecast volatility while preserving epidemiologically meaningful seasonal amplitudes, thereby improving forecast stability and interpretability over extended horizons. Forecasts for 2024--2028 project average monthly malaria admissions ranging from approximately $8{,}000$ to $12{,}200$ cases, consistent with observed seasonal transmission patterns and stable predictive intervals across the forecast horizon.
Spatio-temporal variability analysis additionally revealed pronounced ecological heterogeneity in malaria burden across Ghana, with northern high-burden districts exhibiting comparatively stable relative transmission patterns despite large absolute fluctuations in admissions. These findings highlight the importance of uncertainty-aware forecasting systems capable of supporting adaptive malaria preparedness under heterogeneous transmission conditions.
The proposed framework provides a scalable probabilistic forecasting approach for malaria early warning, seasonal preparedness, and short- to medium-term operational planning in endemic settings. Although constrained by reporting quality and the absence of environmental covariates, this study establishes a methodological foundation for integrating probabilistic machine learning and classical seasonal smoothing within epidemiological forecasting systems designed for resource-constrained public-health environments.
\end{abstract}

\begin{graphicalabstract}
\includegraphics[width=\textwidth]{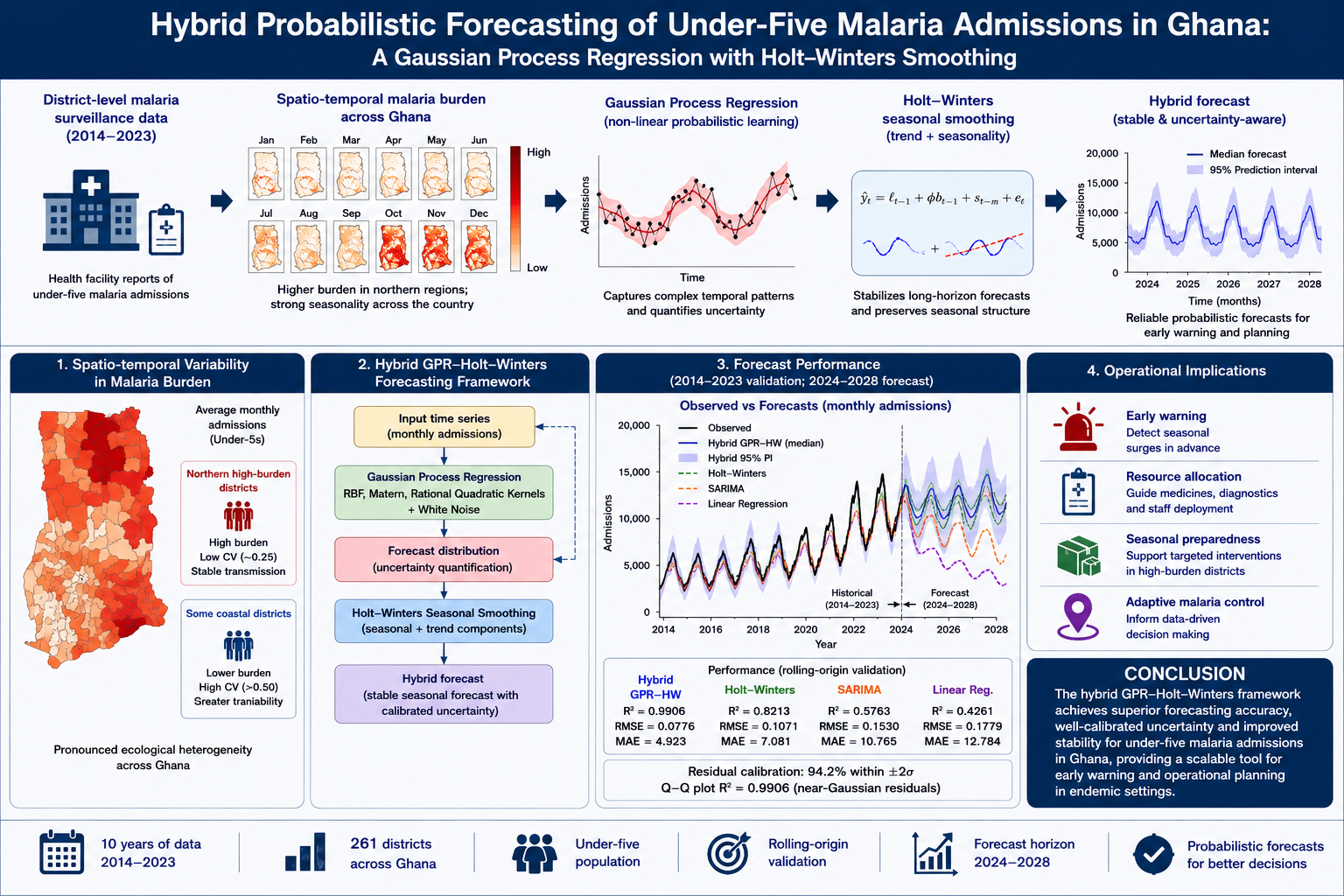}
\end{graphicalabstract}

\begin{highlights}
\item Hybrid GPR–Holt–Winters model enhances malaria forecast accuracy.
\item Probabilistic framework improves uncertainty quantification and stability.
\item Spatial analysis reveals high, stable burdens in northern districts.
\item Robust residual calibration with $94.2 \%$ within $\pm 2\sigma$ limits.
\end{highlights}

\begin{keyword}
Malaria forecasting \sep Gaussian process regression \sep Holt–Winters smoothing \sep Predictive modelling \sep Time-series analysis  \sep Predictive epidemiology



\end{keyword}

\end{frontmatter}



\section{Introduction}
\label{sec1}

Malaria remains a major public health burden across sub-Saharan Africa, driven by persistent challenges such as increasing drug resistance, limited access to healthcare, and fragile health systems \cite{casanova2024artemisinin,oladipo2022increasing,sarpong2022zero}. In 2021 alone, the region accounted for 247 million malaria cases globally, underscoring its disproportionate share of the disease burden \cite{world2022world}. Ghana is among the countries most affected, reporting approximately $5.4$ million malaria cases in $2021$, equivalent to an incidence rate of $164.3$ cases per $1,000$ population and $12,557$ associated deaths\footnote{\url{https://www.afro.who.int/sites/default/files/2023-08/Ghana.pdf}}.

Despite national and global initiatives aimed at malaria control and eventual elimination, the disease remains endemic throughout Ghana. Children under five years of age continue to be the most vulnerable group, reflecting broader regional patterns \cite{aidoo2024malaria,awine2017towards}. Malaria transmission in Ghana is not uniform; it exhibits strong seasonal and spatial variability, with environmental factors such as rainfall, temperature, and humidity influencing transmission peaks across different ecological zones \cite{awine2018spatio,adu2015spatiotemporal}.

To strengthen malaria control efforts, forecasting has become an essential tool for anticipating outbreaks, informing public health planning, and guiding the timely allocation of resources. Accurate and interpretable forecasts at monthly resolution can support early warning systems and localised interventions \cite{myers2000forecasting}. However, developing reliable forecasting tools remains challenging due to the non-linear and non-stationary dynamics of malaria transmission, particularly in regions like Ghana, which has high district-level variability.

Traditional statistical models, such as Autoregressive Integrated Moving Average (ARIMA) and its seasonal extension, SARIMA, have long been used for malaria forecasting due to their simplicity and ease of implementation. However, these models rely on linearity and stationarity assumptions that are often violated in epidemiological time series.
These models are particularly effective for stationary time series with linear dynamics and well-defined seasonal components \cite{box2015time}. Similarly, exponential smoothing methods like Holt-Winters are frequently used for short-term forecasting in health surveillance systems where seasonality is prominent \cite{taylor2003short}. However, these models exhibit several limitations when applied to epidemiological data, which are often characterised by non-linear trends, abrupt regime shifts, and unobserved exogenous influences such as environmental variability and intervention policies \cite{viboud2018rapidd,held2005statistical}.
Moreover, classical time series models do not inherently provide a full probabilistic description of future observations, limiting their ability to express forecast uncertainty in a principled manner. This becomes a critical drawback in public health decision-making, where risk quantification and uncertainty bounds are essential for planning and response.

Recent advances in machine learning and probabilistic modelling have brought powerful tools such as Gaussian Process Regression (GPR) into the spotlight. GPR is a non-parametric Bayesian method capable of learning from limited data, modelling complex seasonality, and providing full predictive distributions rather than point estimates \cite{williams2006gaussian}.
Its ability to incorporate uncertainty directly into forecasts makes it particularly suitable for epidemiological forecasting, where confidence intervals are essential for risk assessment and policy-making \cite{paliwal2023machine}. Nonetheless, the application of GPR to malaria forecasting in African contexts remains underexplored, and its performance relative to classical statistical models is not yet fully understood. More critically, existing studies have largely treated probabilistic machine learning models and classical time-series approaches as competing alternatives rather than complementary components. This has left a specific methodological gap: there is limited work on hybrid frameworks that explicitly address the tendency of GPR to produce increasingly diffuse predictive intervals while preserving the seasonal structure captured by classical methods such as Holt-Winters.

In epidemiological settings characterised by strong seasonality and operational decision constraints, this limitation is non-trivial. Purely probabilistic models may yield forecasts with high uncertainty that are difficult to operationalise, while classical approaches impose structure at the cost of flexibility. A principled integration of these paradigms therefore defines a distinct methodological niche, enabling forecasts that balance uncertainty quantification with seasonal stability.

To address this gap, we propose a comparative study of forecasting models using district-aggregated monthly malaria incidence data from Ghana, covering the period from January 2014 to December 2023. The data include admissions for both children under five and individuals aged five and above, along with spatial coordinates (latitude and longitude) for each district. For this study, we focus primarily on the under-five population due to their heightened vulnerability and high incidence burden. These data are collated from national health surveillance systems and represent a rich temporal dataset amenable to rigorous modelling.

Our study pursues three main objectives. First, we aim to construct forecasting models using four complementary approaches: Linear Regression, Holt-Winters Exponential Smoothing, Seasonal ARIMA (SARIMA), and GPR--selected to represent both classical and probabilistic time-series paradigms. 
Second, we seek to compare the performance of these models in terms of forecasting accuracy and uncertainty quantification over both historical and projected time frames. 
Third, we introduce a hybrid ensemble framework, where GPR forecasts are fine-tuned using Holt-Winters smoothing to mitigate excessively wide confidence intervals, thus improving both interpretability and reliability.
This study contributes to the methodological advancement of malaria incidence forecasting in Ghana by applying GPR to epidemiological time-series data. While GPR has been employed in various health forecasting contexts, its use in modelling malaria trends in sub-Saharan Africa remains relatively limited. By systematically comparing GPR with classical approaches such as SARIMA, Holt-Winters, and linear regression, the study provides a robust assessment of model performance under conditions of seasonality and uncertainty. Furthermore, this study introduces a hybrid forecasting framework that integrates GPR with Holt-Winters smoothing. The motivation arises from a known limitation of probabilistic models, namely the tendency of GPR to produce increasingly wide predictive intervals over extended forecast horizons, which can limit its practical utility in public health planning. 

Within this formulation, Holt-Winters smoothing is applied to the GPR predictive mean, imposing seasonal regularity and improving short-term stability while retaining the uncertainty structure derived from the Gaussian process. The resulting forecasts preserve non-linear pattern learning and probabilistic inference, but exhibit reduced volatility and more interpretable behaviour for decision-making. This approach therefore, moves beyond standard model comparison by offering a structured mechanism for stabilising probabilistic forecasts in epidemiological time series, particularly in settings where seasonal consistency and uncertainty awareness must be balanced.

To complement the forecasting framework and contextualise its outputs, the study incorporates additional analytical measures that support interpretation of malaria dynamics at the district level. The coefficient of variation (CV) is used to quantify temporal instability in monthly admissions, providing a scale-independent measure of variability that complements the model-based forecasts. This enables a clear distinction between districts with persistently high but stable burden and those with lower yet highly variable transmission, linking predictive results to operational planning needs. The use of ten years of nationwide, district-aggregated data further situates these temporal patterns within Ghana’s ecological and spatial context, highlighting contrasts between relatively stable high-burden northern districts and more variable low-burden areas elsewhere.

The analysis is conducted within a univariate forecasting framework based on routinely reported malaria admissions. This choice reflects the structure of Ghana’s national surveillance system, where consistent district-level case counts are available, but temporally aligned environmental and intervention covariates are not systematically recorded across all districts for the 2014--2023 period. Establishing this baseline ensures methodological consistency and direct applicability to operational data streams, while providing a foundation for future extensions that incorporate climate and intervention variables as they become available.

The remainder of this paper is organised as follows. Section~\ref{sec:studyarea} introduces the study area, describing the geographical setting of Ghana, the epidemiological characteristics of malaria transmission, and the surveillance datasets used in the analysis. Section~\ref{sec:method} presents the methodological framework, detailing the classical statistical models, the proposed hybrid GPR--Holt--Winters forecasting architecture, and the procedures adopted for calibration, uncertainty estimation, and rolling-origin validation. Section~\ref{sec:RNA} presents the results and analysis, including comparative forecasting performance, residual diagnostics, uncertainty behaviour, and spatio-temporal variability in malaria admissions across ecological zones. Section~\ref{sec:disc} provides a broader interpretation of the findings within the context of probabilistic epidemiological forecasting, discussing methodological implications, model robustness, operational relevance, and limitations of the current framework. Finally, Section~\ref{sec:conc} summarises the principal contributions of the study, highlights its implications for adaptive malaria surveillance and public-health planning, and outlines directions for future methodological and operational extensions.

\section{Study Area} \label{sec:studyarea}

\begin{figure*}
\begin{minipage}[H]{\linewidth}
\centering
\includegraphics[width=\textwidth]{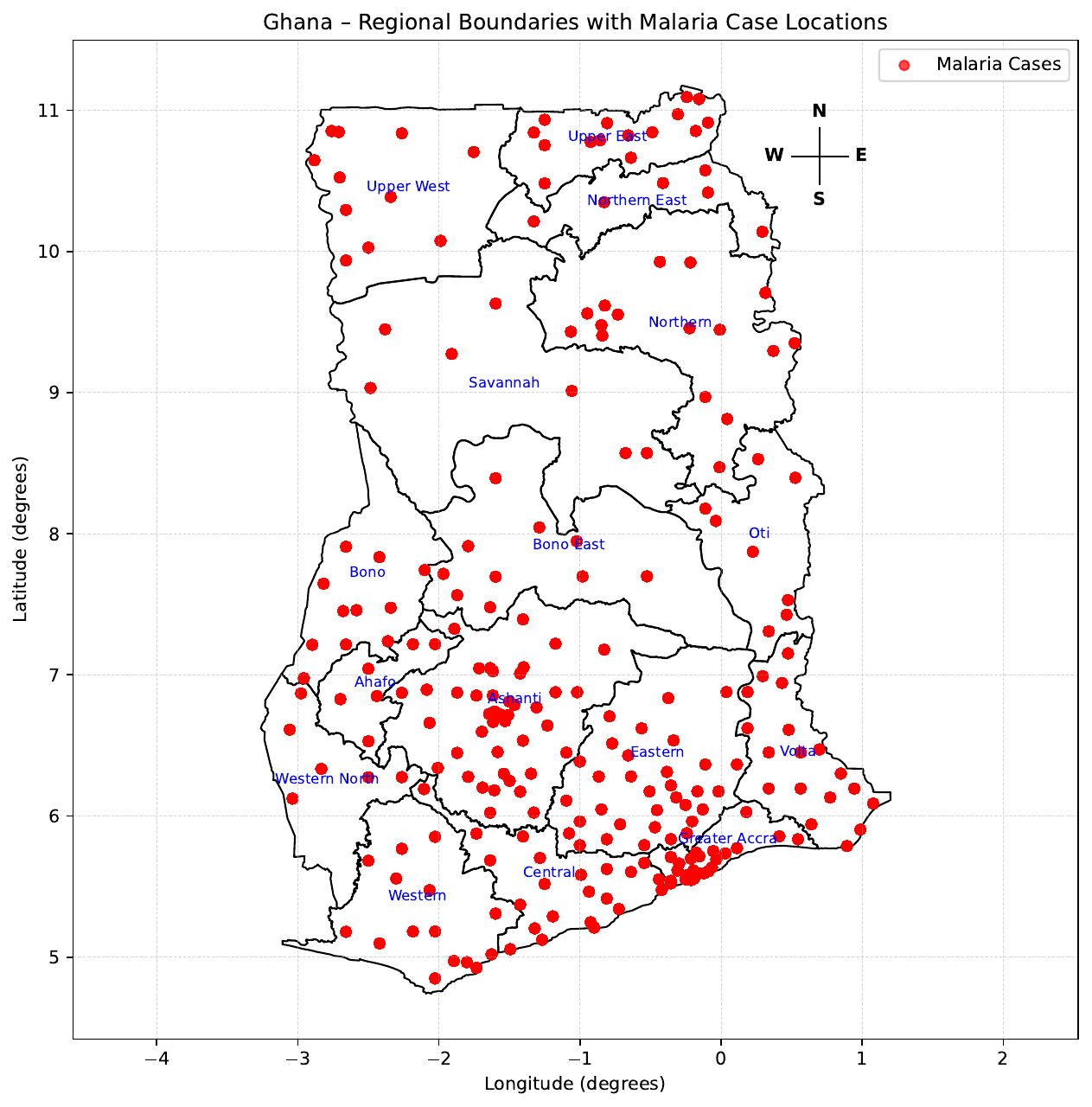} 
\end{minipage}
\caption{Map of Ghana showing regional boundaries and the geospatial distribution of malaria case reporting sites. Red circles represent malaria surveillance locations aggregated at the district level, overlaid on the administrative boundaries of the sixteen regions of Ghana. Regional names are annotated in blue for reference.
}\label{fig:ghmap_f1}
\end{figure*}

Ghana, located along the Gulf of Guinea in West Africa, lies between latitudes 4.5\textdegree N and 11.5\textdegree N and longitudes 3.5\textdegree W and 1.5\textdegree E. It covers an area of approximately $239,460\, \text{km}^2$, including $8,520 \, \text{km}^2$ of water bodies and is administratively divided into sixteen regions and $261$ districts.
The study focuses on malaria case surveillance across all regions, with spatial aggregation at the district level.
Fig.~\ref{fig:ghmap_f1} illustrates the regional boundaries of Ghana along with the geographic distribution of reported malaria cases used in this analysis. The dataset includes observations from both northern and southern ecological zones, providing temporal records suitable for time series modelling.
The spatial distribution of the cases supports national-level modelling while preserving sub-national variability essential for understanding local malaria trends.

Ghana’s ecological and climatic diversity contributes significantly to spatial heterogeneity in malaria transmission. 
The country is typically grouped into three broad ecological belts: the Guinea Savannah in the north, the Forest zone in the centre, and the Coastal Savannah in the south \cite{adu2015spatiotemporal,de2010environmental}.
These zones experience distinct rainfall patterns and hydrological profiles, which shape the seasonality and intensity of malaria transmission. The northern regions generally experience a unimodal rainy season (May--October), while the middle and southern belts have bimodal rainfall peaks (April--June and September--November), influencing vector breeding and transmission cycles accordingly \cite{awine2018spatio}.

Malaria remains endemic throughout Ghana, with a consistently high incidence rate reported among children under five years of age, the most vulnerable demographic group \cite{adokiya2025perspectives,aidoo2024malaria,kolekang2022challenges}.
Despite ongoing control efforts, the disease exhibits pronounced spatial and seasonal variations, reflecting disparities in vector ecology, healthcare accessibility, and socioeconomic factors across regions. Urban areas, particularly Greater Accra, may report lower prevalence due to improved housing, drainage, and health infrastructure, whereas higher transmission persists in rural districts with limited access to preventive and curative services \cite{fobil2011neighborhood}.

\section{Methodology} \label{sec:method}

\subsection{Data Description}
\subsubsection{Spatial Characteristics of Malaria Admissions Data} \label{sub:spatial}

The dataset utilised in this study comprises district-level records of malaria admissions among children under five years across Ghana, spanning a ten-year period (2014--2023).
Each record includes the number of monthly admissions, the associated district, and the geographical coordinates (latitude and longitude) of reporting centres or centroids. 
These data were sourced from the Ghana Health Service's facility-based reports and represent cumulative counts for each calendar month.

\begin{figure*}
\begin{minipage}[H]{\linewidth}
\centering
\includegraphics[width=\textwidth]{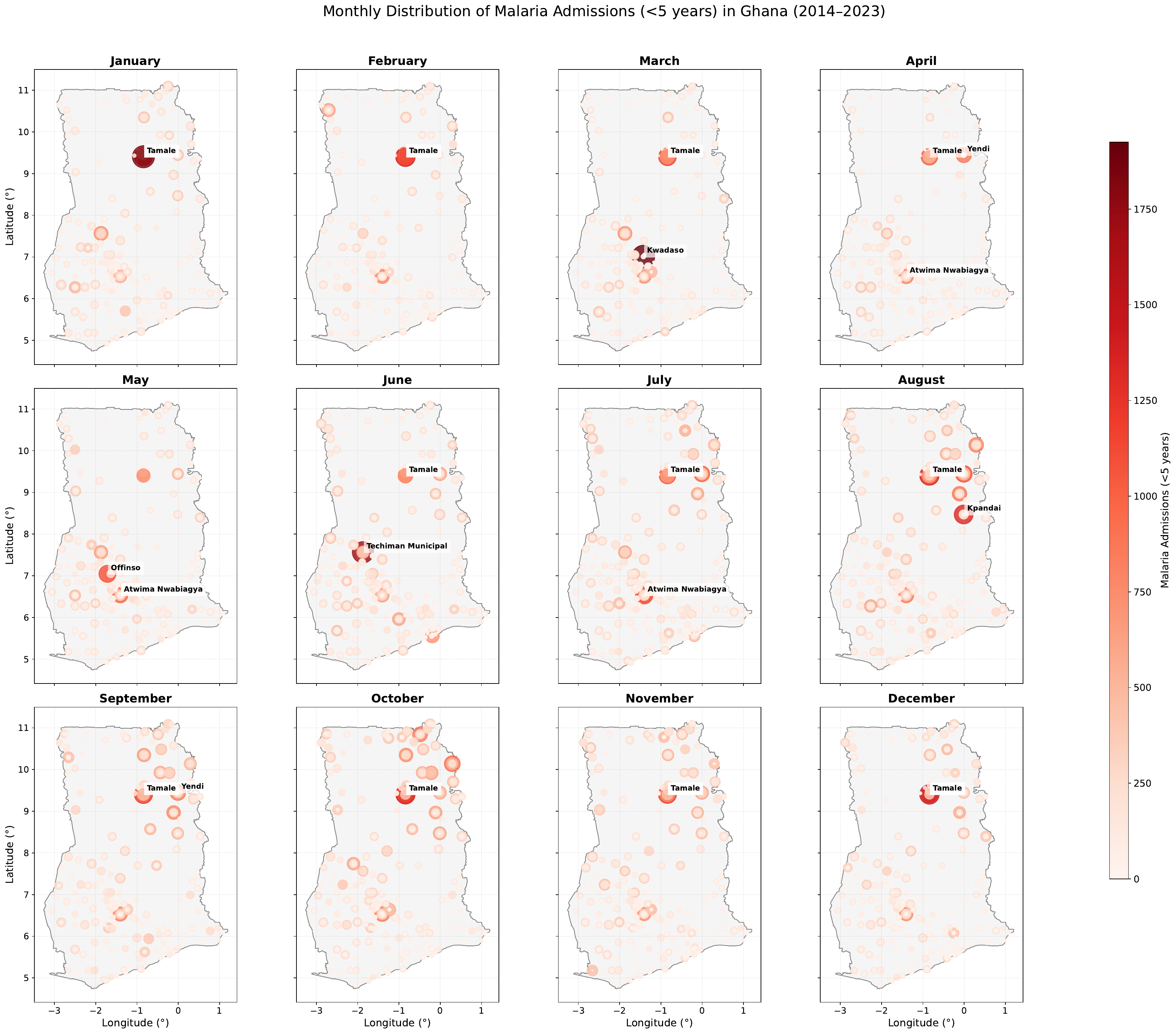} 
\end{minipage}
\caption{Monthly spatial distribution of under-five malaria admissions across Ghana (2014--2023). Each panel represents a cumulative visualisation of reported cases in a given month aggregated over ten years. Marker size and colour intensity are proportional to the number of admissions, highlighting persistent and seasonal hotspots. Tamale emerges as the most consistent high-burden district across the year, particularly in the wet season months (April--October).
}\label{fig:monthly_malaria_hotspots_f2}
\end{figure*}

To assess the spatial footprint of the malaria burden, the data were aggregated by district and month, and subsequently converted to a geospatial format using district centroids. Fig.~\ref{fig:monthly_malaria_hotspots_f2} presents the monthly spatial distribution of under-five malaria admissions in different districts.
Each district is represented by a marker whose size and colour intensity reflect the magnitude of admissions in a given month. This spatial rendering enables the examination of geographic variability and seasonal dynamics simultaneously.

Preliminary visual analysis reveals pronounced spatial heterogeneity. Northern districts such as Tamale Metropolitan, which consistently appears as the district with the highest monthly burden, recorded peak admissions exceeding $1,800$ cases per month (notably in January, June, and December). Other high-burden districts include Techiman Municipal (notably in June) and Atwima Nwabiagya (April–July). By contrast, districts in Greater Accra (e.g., Klottey-Korle) and parts of the Central Region showed relatively low admissions, often fewer than $200$ monthly cases, throughout the year.

These patterns align with Ghana’s ecological zones, where the Guinea and Sudan savannah belts in the north experience higher malaria transmission intensity, while the coastal savannah and forest zones in the south benefit from better drainage, urbanisation, and health infrastructure, all of which suppress malaria transmission \cite{peprah2024patient,oheneba2022estimating,asante2011malaria, asante2003economic}.

Routine surveillance data may be influenced by changes in reporting practices, diagnostic capacity, and data completeness over time. During the 2014--2023 period, Ghana’s malaria surveillance system experienced gradual improvements in reporting coverage and data digitisation, particularly through the expansion of the District Health Information Management System (DHIMS2). These changes may have introduced variation in reported case counts that is not solely attributable to underlying transmission dynamics.

No major nationwide reporting disruptions were identified over the study period; however, gradual improvements in data quality may introduce mild bias in observed trends. This limitation is inherent to routine surveillance data and is taken into account when interpreting long-term patterns and model outputs.

\subsubsection{Temporal instability of malaria burden} \label{sub:temporal}

Malaria transmission in Ghana is highly heterogeneous, with seasonal peaks and year-to-year variability differing across ecological zones. Such fluctuations influence service demand, treatment provision, and the reliability of incidence forecasts. To capture this instability in a way that is comparable across districts with different mean burdens, we employed the coefficient of variation (CV), a scale-independent measure of relative dispersion \citep{Reed2002}. 

For each district $i$, with monthly malaria admissions $x_{i,t}$ observed over $N$ months, the CV was defined as:

\begin{equation}
\mathrm{CV}_i = \frac{\sigma_i}{\mu_i} \,,
\label{eq:cv}
\end{equation}

\noindent where $\mu_i$ is the mean and $\sigma_i$ the standard deviation of monthly admissions. A higher CV indicates greater relative instability in the temporal pattern of malaria cases, whereas a lower CV reflects more consistent transmission dynamics. 
This approach is particularly suited to the Ghanaian context, where districts with similar ecological profiles (for example, northern savannah, forest, and coastal belts) may exhibit very different temporal signatures of malaria admissions despite comparable annual totals. 
The CV provides a measure of relative variability. This enables identification of districts where forecasting uncertainty is driven by unstable transmission dynamics, as opposed to those where malaria trends are more predictable.

In addition to the CV, we also report the standard deviation $\sigma_i$, which quantifies the absolute magnitude of fluctuations in case counts. While CV highlights relative instability, $\sigma_i$ reflects the scale of variation that directly impacts health system operations, such as bed occupancy, medicine stock-outs, and staff allocation. Reporting both measures, therefore, provides complementary insights: CV identifies unstable temporal dynamics irrespective of burden, whereas $\sigma_i$ captures fluctuations of operational significance. 

\subsubsection{Traditional forecasting approaches} \label{sub:traditional_forecasting}

To establish a performance baseline and assess the suitability of classical time-series models for malaria incidence forecasting, three well-established approaches were implemented: simple linear regression, Holt–Winters exponential smoothing, and the Seasonal Autoregressive Integrated Moving Average (SARIMA) model. These models were selected because of their interpretability, computational efficiency, and long-standing use in epidemiological forecasting, particularly for diseases exhibiting both secular trends and seasonal dynamics \citep{akter2024comparative,MABASO2007}.
The analyses were conducted on monthly malaria admission data for children under five years of age, covering the period 2014--2023. The forecasting horizon was set to sixty months (2024--2028), consistent to assess short- to medium-term predictive performance across varying transmission regimes.

\paragraph{\normalfont \emph{Linear regression model}:}
A baseline deterministic model was first fitted using ordinary least squares linear regression, where monthly malaria admissions $y_t$ were expressed as a function of time $t$:
\begin{equation}
y_t = \beta_0 + \beta_1 t + \varepsilon_t ,
\label{eq:linear_model}
\end{equation}
\noindent where $\beta_0$ and $\beta_1$ denote the intercept and slope coefficients, respectively, and $\varepsilon_t$ represents the residual error term assumed to follow $\varepsilon_t \sim \mathcal{N}(0,\sigma^2)$. The fitted trend was extrapolated for the forecast period, and $95\%$ confidence intervals were derived under the assumption of homoscedastic residual variance. 
Linear regression captures long-term trends but does not model seasonality or residual dependence, and therefore serves primarily as a comparative benchmark rather than a fully dynamic forecasting tool.

\paragraph{\normalfont \emph{Holt-Winters exponential smoothing}:}
To capture both level and seasonal components, Holt–Winters exponential smoothing was employed using an additive formulation. The method estimates the smoothed level $L_t$, trend $T_t$, and seasonal component $S_t$ through the following recursive equations:
\begin{align}
L_t &= \alpha (y_t - S_{t-s}) + (1-\alpha)(L_{t-1} + T_{t-1}), \label{eq:hw_level} \\
T_t &= \beta (L_t - L_{t-1}) + (1-\beta)T_{t-1}, \label{eq:hw_trend} \\
S_t &= \gamma (y_t - L_t) + (1-\gamma)S_{t-s}, \label{eq:hw_seasonal}
\end{align}
\noindent where $\alpha$, $\beta$, and $\gamma$ are smoothing parameters in $(0,1)$, and $s=12$ denotes the seasonal period corresponding to one year. Forecasts for $h$ steps ahead are generated as:
\begin{equation}
\hat{y}_{t+h} = L_t + hT_t + S_{t-s+h_s},
\label{eq:hw_forecast}
\end{equation}
\noindent where $h_s = ((h-1) \bmod s) + 1$. The Holt–Winters method captures both trend and seasonal components in malaria time series, reflecting underlying epidemiological dynamics \citep{assefa2025relationship,abiodun2018exploring}.

\paragraph{\normalfont \emph{SARIMA}:}
To explicitly model autocorrelation and stochastic seasonality, a SARIMA$(p,d,q)(P,D,Q)_s$ process was fitted to the monthly series. The general specification is given by:
\begin{equation}
\Phi_P(B^s)\phi_p(B)(1-B)^d(1-B^s)^D y_t = \Theta_Q(B^s)\theta_q(B)\varepsilon_t ,
\label{eq:sarima}
\end{equation}
\noindent where $B$ is the backshift operator ($B y_t = y_{t-1}$), $\phi_p(B)$ and $\theta_q(B)$ represent the non-seasonal autoregressive (AR) and moving average (MA) polynomials, respectively, and $\Phi_P(B^s)$ and $\Theta_Q(B^s)$ denote their seasonal counterparts. The parameters $(p,d,q)$ and $(P,D,Q)$ indicate the non-seasonal and seasonal orders, and $s=12$ represents the annual seasonal cycle. In this study, the selected model was SARIMA$(1,1,1)(1,1,1)_{12}$, chosen based on autocorrelation and partial autocorrelation diagnostics and minimisation of the Akaike Information Criterion (AIC). SARIMA has been widely applied in modelling malaria and other vector-borne diseases in sub-Saharan Africa, capturing both inter-annual persistence and intra-annual seasonality in incidence data \citep{armando2025spatio,ebhuoma2018seasonal}.

\paragraph{\normalfont \emph{Model evaluation and comparison}:}
Model performance was evaluated using standard forecast accuracy and diagnostic statistics. The root mean square error (RMSE) and mean absolute error (MAE) were computed to assess average forecast deviation, while the coefficient of determination ($R^2$) quantified the proportion of explained variance. Information-theoretic criteria, namely the Akaike Information Criterion (AIC) and Bayesian Information Criterion (BIC), were used to balance model fit against complexity. Residuals were further tested for normality (Shapiro–Wilk, Jarque–Bera, and Kolmogorov–Smirnov tests) and for skewness and kurtosis departures from Gaussianity. Stationarity was assessed using the augmented Dickey–Fuller (ADF) test, and the residual autocorrelation structure was inspected through the autocorrelation and partial autocorrelation functions (ACF and PACF). These diagnostics collectively ensured that model assumptions were satisfied and that forecast uncertainty was appropriately characterised.

The comparative analysis of the three conventional forecasting models--linear regression, Holt--Winters, and SARIMA, provides an empirical basis for identifying which statistical approach most effectively captures the long-term trend, seasonal structure, and temporal dependence of malaria admissions in Ghana. Such benchmarking is essential for establishing a performance baseline against which more advanced forecasting frameworks, including probabilistic and hybrid ensemble approaches, can be rigorously evaluated.

In this context, Artificial Neural Networks (ANNs) and related deep-learning models were not included in this study because they typically require large and diverse training samples to achieve stable parameter estimation and avoid overfitting. The dataset analysed here consists of a single uninterrupted monthly time series of $120$ observations, which does not provide the sample volume necessary for reliable neural-network training, hyperparameter optimisation, or independent validation. In addition, ANN-based approaches generally do not provide analytically interpretable uncertainty estimates, which are essential for public-health forecasting, risk communication, and operational planning.

For these reasons, the comparative framework focuses on models that are statistically well-defined for short epidemiological time series and that provide transparent, uncertainty-aware outputs suitable for operational decision-making. Classical approaches such as Holt--Winters and SARIMA offer interpretable representations of trend, seasonality, and temporal dependence while remaining computationally stable under limited-data conditions. Gaussian Process Regression (GPR), although more flexible, remains well-suited to small-sample forecasting because it operates within a Bayesian probabilistic framework in which predictive uncertainty is estimated directly through the posterior covariance structure. This enables simultaneous quantification of both expected disease burden and forecast confidence intervals. Such uncertainty-aware forecasting is particularly important in malaria surveillance systems, where predictive models must support intervention timing, resource allocation, and outbreak preparedness under highly variable transmission conditions.

\subsection{Hybrid Gaussian Process–Holt–Winters Forecasting Framework}
\label{sub:gpr_hybrid}

The GPR forms the probabilistic foundation of the forecasting system and is combined with Holt–Winters (HW) smoothing to enhance short-term stability. This hybrid formulation is designed to accommodate the non-linear and seasonally structured behaviour of malaria incidence, to quantify uncertainty through a principled Bayesian framework, and to produce forecasts that remain operationally coherent at extended lead times.

Given monthly time indices $X=[x_1,\ldots,x_n]^\top$ and observed case counts $\mathbf{y}=[y_1,\ldots,y_n]^\top$, the latent process is modelled using the Gaussian process prior
\begin{equation}
    f(\mathbf{x}) \sim \mathcal{GP}(0, k(\mathbf{x},\mathbf{x}')),
    \label{eq:gpr_prior2}
\end{equation}
which induces a closed-form posterior at new inputs $X_*$. The mean and covariance of this posterior are given by
\begin{equation}
\boldsymbol{\mu}_* = K_*^\top (K + \sigma_n^2 I)^{-1}\mathbf{y},
\label{eq:gpr_mean2}
\end{equation}
\begin{equation}
\Sigma_* = K_{**} - K_*^\top (K + \sigma_n^2 I)^{-1}K_*,
\label{eq:gpr_cov2}
\end{equation}
where $K$, $K_*$ and $K_{**}$ denote the relevant kernel matrices.  
Equation~\eqref{eq:gpr_mean2} represents the model’s expected continuation of the malaria time series, while the diagonal of Eq.~\eqref{eq:gpr_cov2} provides horizon-specific uncertainty estimates. The explicit linkage between forecast uncertainty and the structure of the covariance kernel is critical: months that historically display volatile dynamics or irregular reporting naturally produce wider credible intervals.

The composite kernel adopted in this study,
\begin{equation}
k(\cdot,\cdot)=
k_{\mathrm{SE}} +
k_{\mathrm{Per}} +
k_{\mathrm{RQ}} +
k_{\mathrm{Long}} +
k_{\mathrm{WN}},
\label{eq:kernel2}
\end{equation}
allows the GP to represent multiple epidemiologically meaningful scales simultaneously. The squared-exponential term captures smooth intra-seasonal variation, the periodic component encodes the dominant annual cycle, the rational quadratic term accounts for medium-scale non-stationarity, the long-term kernel accommodates gradual structural change, and the white-noise term models reporting variability. Through the additive form of Eq.~\eqref{eq:kernel2}, these components contribute independently to the covariance structure, enabling the GP posterior to capture complex departures from idealised seasonal patterns without explicit manual decomposition. Hyperparameters were estimated via marginal-likelihood optimisation with multiple restarts, subject to biologically plausible constraints such as limiting the periodic component to 6--24 months.

The hybrid integration with Holt–Winters builds directly on the statistical properties of the GP posterior. The mean function in Eq.~\eqref{eq:gpr_mean2} captures smooth deviations from seasonal structure, while Holt–Winters provides a stable seasonal backbone that extrapolates reliably at long lead times. This complementarity leads to two natural integration pathways.

The first approach augments the GP inputs with Holt–Winters fitted values $h_t^{\mathrm{HW}}$, giving the decomposition
\begin{equation}
    y_t = h_t^{\mathrm{HW}} + r_t, \qquad r_t \sim \mathcal{GP}(0, k_r),
    \label{eq:residual_gp}
\end{equation}
where the GP focuses on modelling the residual component $r_t$. In this configuration, Eq.~\eqref{eq:gpr_mean2} yields a probabilistic correction to the HW seasonal pattern, and Eq.~\eqref{eq:gpr_cov2} supplies corresponding uncertainty estimates.

The second integration pathway smooths the GP predictive mean trajectory using Holt–Winters,
\begin{equation}
    \mu_{\mathrm{hyb},t+h} = S\!\big(\mu_{\mathrm{GP},t+h}\big),
    \label{eq:hw_smoothing}
\end{equation}
and adjusts the associated predictive variance through
\begin{equation}
    \sigma^2_{\mathrm{hyb},t+h} = \sigma^2_{\mathrm{GP},t+h} + \sigma^2_{\mathrm{res}},
    \label{eq:var_propagation}
\end{equation}
where $\sigma^2_{\mathrm{GP},t+h}$ corresponds to the diagonal elements of Eq.~\eqref{eq:gpr_cov2}.  
Equations~\eqref{eq:hw_smoothing}–\eqref{eq:var_propagation} ensure that the hybrid forecast preserves the probabilistic information generated by the GP while benefiting from the structural regularity imposed by HW.

The posterior expressions in Eqs.~\eqref{eq:gpr_mean2}–\eqref{eq:gpr_cov2} require inversion of $(K+\sigma_n^2 I)$, leading to a computational cost of $O(n^3)$ and memory usage of $O(n^2)$. With approximately 120 monthly observations per district, these requirements are modest, and training completes within seconds. Holt–Winters adds negligible overhead. A detailed assessment of runtime behaviour and potential extensions for real-time deployment is provided in Appendix~\ref{supp:complexity}.

Figure~\ref{fig:hybrid_workflow} summarises the complete hybrid modelling pipeline, linking standardisation, GPR fitting, posterior computation through Eqs.~\eqref{eq:gpr_mean2}--\eqref{eq:gpr_cov2}, multi-step forecasting, and seasonal smoothing via Eq.~\eqref{eq:hw_smoothing}. This schematic highlights how each component contributes to the construction of the final hybrid forecast.

The final forecasting results presented in this study were generated using the hybrid GPR--Holt--Winters pathway illustrated in Fig.~\ref{fig:hybrid_workflow}. In this configuration, GPR was first employed to model the non-linear temporal structure of the malaria time series and to estimate predictive uncertainty through the posterior mean and covariance functions. The resulting forecasts were subsequently refined using Holt--Winters seasonal smoothing to preserve recurrent seasonal behaviour and reduce short-term volatility in the long-horizon projections. This sequential integration was adopted to combine the probabilistic flexibility of GPR with the seasonal stability and interpretability of exponential smoothing, thereby producing forecasts that remain both uncertainty-aware and operationally robust under limited-data conditions.
The selected pathway therefore provides a unified forecasting framework capable of representing non-linear transmission behaviour while maintaining stable seasonal structure and interpretable uncertainty estimates. Such characteristics are particularly important for malaria surveillance applications, where forecasting systems must remain reliable under limited-data conditions and variable transmission dynamics.

When implemented across every district, the hybrid framework produces 60‑month forecasts whose predictive intervals are well calibrated. These outputs support estimation of expected and upper-bound case loads, facilitate identification of districts with elevated forecast uncertainty, and enhance planning for case-management capacity and seasonal preparedness. The framework therefore combines the interpretability and seasonal stability of Holt--Winters with the uncertainty-aware flexibility of GPR, providing a forecasting system well suited to the heterogeneous and seasonally driven malaria transmission dynamics in Ghana.

\begin{figure}
\begin{minipage}[H]{\linewidth}
\centering
\includegraphics[width=0.9\textwidth]{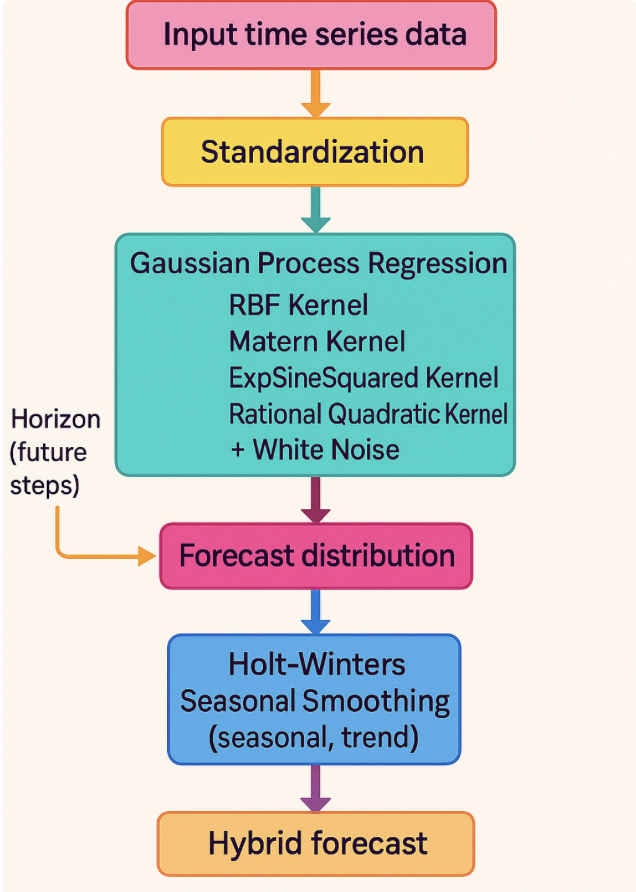}
\end{minipage}
\caption{
Workflow of the hybrid GPR-Holt–Winters forecasting framework.
The diagram illustrates the modelling stages, beginning with standardisation and composite-kernel GPR training, followed by computation of the posterior mean and covariance (Eqs.~\eqref{eq:gpr_mean2} and \eqref{eq:gpr_cov2}), multi-step forecasting, and seasonal smoothing using Eq.~\eqref{eq:hw_smoothing}. The hybrid output retains the probabilistic structure of the GPR while adopting the seasonal coherence of Holt–Winters.
}
\label{fig:hybrid_workflow}
\end{figure}

Given the relatively limited length of the time series, a conventional static train--test split (e.g., 80/20) was not adopted, as such partitioning can reduce statistical stability and limit the ability of seasonal models to learn recurring temporal structure. Instead, model performance was evaluated using a rolling-origin (expanding window) time-series validation scheme with a fixed 12-month forecast horizon. At each iteration, models were trained on all observations available up to a given forecast origin and evaluated on the subsequent unseen 12-month period. This procedure was repeated sequentially across the dataset, generating multiple out-of-sample forecast windows spanning 2019--2023.

Performance metrics described in Section~\ref{sub:traditional_forecasting}, including RMSE, MAE, RMSLE, and coefficient of determination ($R^2$), were computed for each validation fold and aggregated using the median across folds to provide a robust estimate of typical predictive accuracy and reduce sensitivity to anomalous forecast windows.

To assess whether observed differences in predictive performance between competing models were statistically significant, pairwise Diebold--Mariano (DM) tests were conducted using squared forecast errors derived from the rolling-origin validation procedure. The DM test evaluates the null hypothesis that two competing forecasting models possess equal predictive accuracy.

For completeness, in-sample model fits were retained solely for diagnostic purposes, including residual analysis and visual inspection of model behaviour. These results are presented separately and are not used to assess predictive performance.

\section{Results and Analysis}\label{sec:RNA}

\begin{figure*}
\begin{minipage}[H]{\linewidth}
\centering
\includegraphics[width=0.9\textwidth]{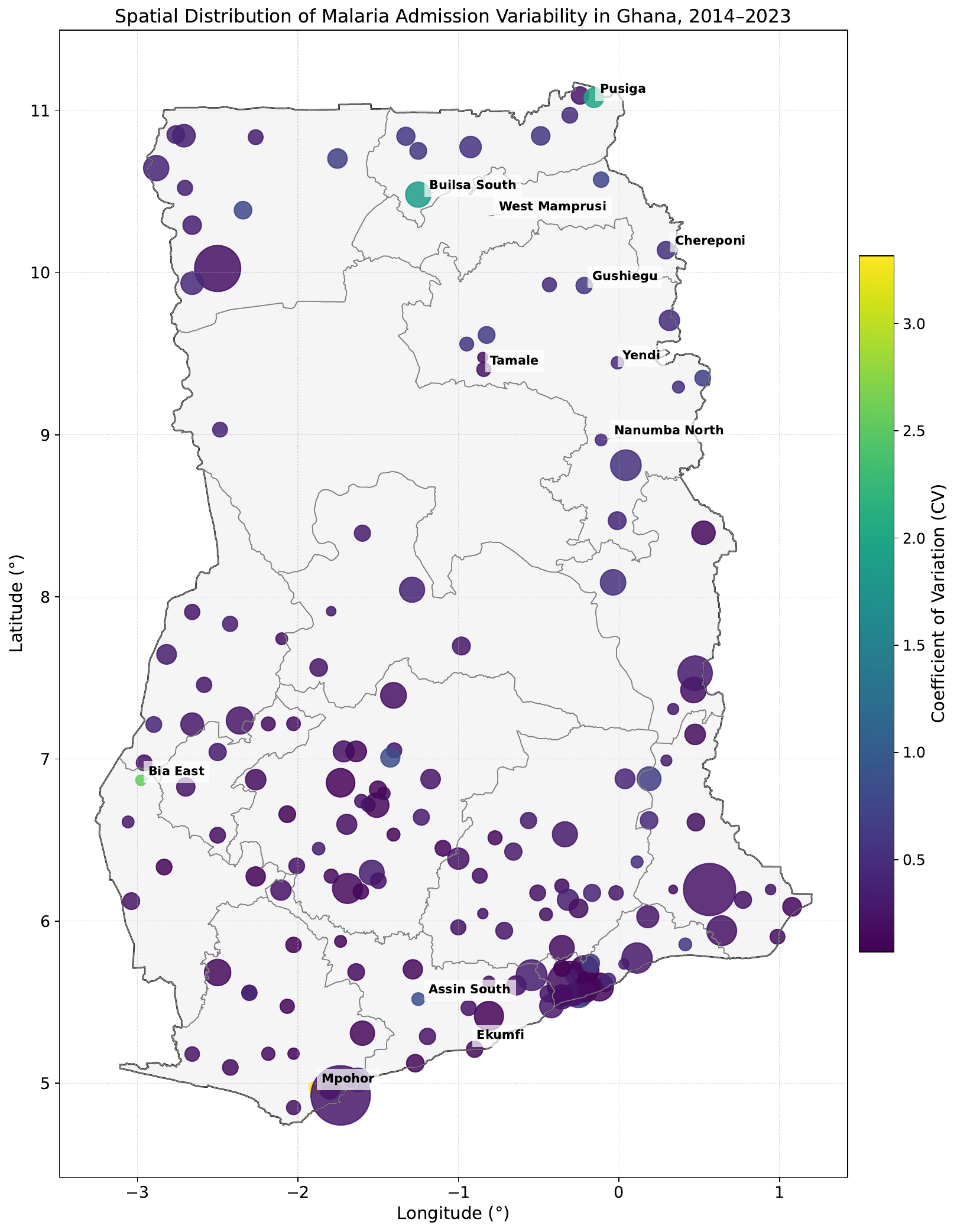} 
\end{minipage}
\caption{Spatial distribution of the coefficient of variation in monthly malaria admissions among children under five. Marker size and colour intensity indicate the magnitude of CV for each district. Higher values reflect greater month-to-month variability in malaria burden.
}\label{fig:CV_malaria_hotspots_f3}
\end{figure*}

The distribution of the CV across districts revealed marked spatial heterogeneity in the stability of monthly malaria admissions (Fig.~\ref{fig:CV_malaria_hotspots_f3}).
Across the country, CV values ranged from $0.12$ to $3.32$, with a national median of $0.61$.
This indicates that, in at least half of Ghana’s districts, month-to-month variation in malaria admissions exceeded $60\%$ of the local mean, underscoring the challenge of achieving consistent predictability in routine service demand. 

Districts with the highest relative variability were typically low-burden settings.
Mpohor, for example, recorded the largest CV ($3.32$) despite having an average of only $0.5$ monthly admissions. 
Similarly, Bia East ($\rm CV = 2.55$) and Pusiga ($\rm CV = 1.99$) exhibited strong fluctuations relative to their means ($17$ and $19$ monthly admissions, respectively). These results highlight the interpretative limitation of CV when case numbers are small: apparent “instability” may reflect low denominators rather than meaningful epidemiological volatility. 

By contrast, districts with very high case volumes displayed more stable relative patterns, but large fluctuations in absolute terms. Tamale Metropolitan, which averaged over $6,800$ monthly admissions, recorded a low CV ($0.25$), yet experienced the largest absolute variability with a standard deviation exceeding $1,600 $cases.
Other high-burden northern districts such as Yendi, Chereponi, and Gushiegu showed moderate CV values ($0.47$--$0.61$) but ranked among the highest in terms of absolute month-to-month fluctuations (standard deviations between $900$ and $1,200$ cases; Table~\ref{tab:variability}).
These findings are operationally important, as even relatively “stable” districts in a statistical sense may generate large swings in patient load when the mean burden is high.

The geographic distribution of instability revealed clear ecological contrasts.
In the northern savannah belt, where malaria transmission is strongly seasonal and peaks after the rains, districts such as Yendi, Chereponi, and Nanumba North combined moderate CV values with large absolute fluctuations. These dynamics mirror the climatic drivers of transmission and reflect the higher risk of service bottlenecks during seasonal surges. In the coastal zone, by contrast, districts such as Assin South ($\rm CV = 0.89$) and Ekumfi ($\rm CV = 1.11$) exhibited comparable or higher CV values but at far lower case volumes, producing fluctuations of limited operational consequence. Urban centres such as Tamale Metropolitan and, to a lesser extent, Yendi stand out as high-burden hubs where fluctuations, though proportionally smaller, translate into significant absolute demand. 

Figure~\ref{fig:CV_malaria_hotspots_f3} makes these patterns visible. Circle colour corresponds to CV magnitude (yellow indicating higher variability, green greater stability), while circle size scales with CV. To avoid over-annotation and facilitate interpretation, twelve districts were labelled: the six with the largest CV values ($0.89$--$3.32$) and the six with the largest standard deviations ($839$--$1684$). 
This dual labelling illustrates both categories of concern: extreme relative variability in low-burden settings, and large absolute fluctuations in high-burden districts. Together, they demonstrate that temporal instability is not randomly distributed but aligns with ecological and demographic gradients, particularly between the northern savannah and southern coastal belts.

These results have important implications for malaria forecasting and health-system preparedness in Ghana. Districts with high CV but low burden (for example, Mpohor, Bia East) may not represent priority zones for resource allocation, but their instability signals the need for caution when generating sub-district forecasts from sparse data. Conversely, high-burden districts with moderate CV but very large standard deviations (e.g., Tamale, Yendi, Chereponi) represent the greatest operational risk: here, even small relative shifts in incidence equate to hundreds of additional admissions in a given month. As shown in Table~\ref{tab:variability}, these districts dominate the ranking of absolute variability and therefore merit special attention in both forecasting model design and contingency planning. 

Taken together, the CV and the standard deviation provide complementary insights into temporal instability. The CV is most useful for identifying unstable trends relative to burden, while the standard deviation highlights the absolute scale of fluctuations that must be managed by health services. The presentation of both measures ensures that decision-making is not skewed by either small denominators in low-burden districts or by the apparent stability of high-burden districts whose fluctuations remain substantial in operational terms. In the Ghanaian context, this dual perspective is essential for forecasting frameworks intended to anticipate case surges and support resource allocation across diverse ecological zones.

\begin{table}[ht] \centering \caption{Districts with the highest CV and variability--burden index ($P=\mu\times CV$). Values are derived from monthly malaria admissions (2014--2023).} \label{tab:variability} \begin{tabular}{lccc} \hline \textbf{District} & \textbf{Mean admissions} & \textbf{CV} & \textbf{Index $P$} \\ \hline Mpohor & 0.5 & 3.32 & 1.66 \\ Bia East & 17.3 & 2.55 & 44.0 \\ Pusiga & 19.4 & 1.99 & 38.7 \\ Builsa South & 13.8 & 1.88 & 25.9 \\ Ekumfi & 8.3 & 1.11 & 9.3 \\ Assin South & 10.4 & 0.89 & 9.3 \\ Tamale Metro & 6860.3 & 0.25 & 1683.7 \\ Yendi & 2643.8 & 0.47 & 1240.0 \\ Chereponi & 1842.6 & 0.52 & 960.4 \\ Gushiegu & 1527.9 & 0.61 & 929.5 \\ Nanumba North & 1940.6 & 0.46 & 884.3 \\ West Mamprusi & 1977.3 & 0.42 & 839.3 \\ \hline \end{tabular} \end{table}

\begin{figure*}
\begin{minipage}[H]{\linewidth}
\centering
\includegraphics[width=\textwidth]{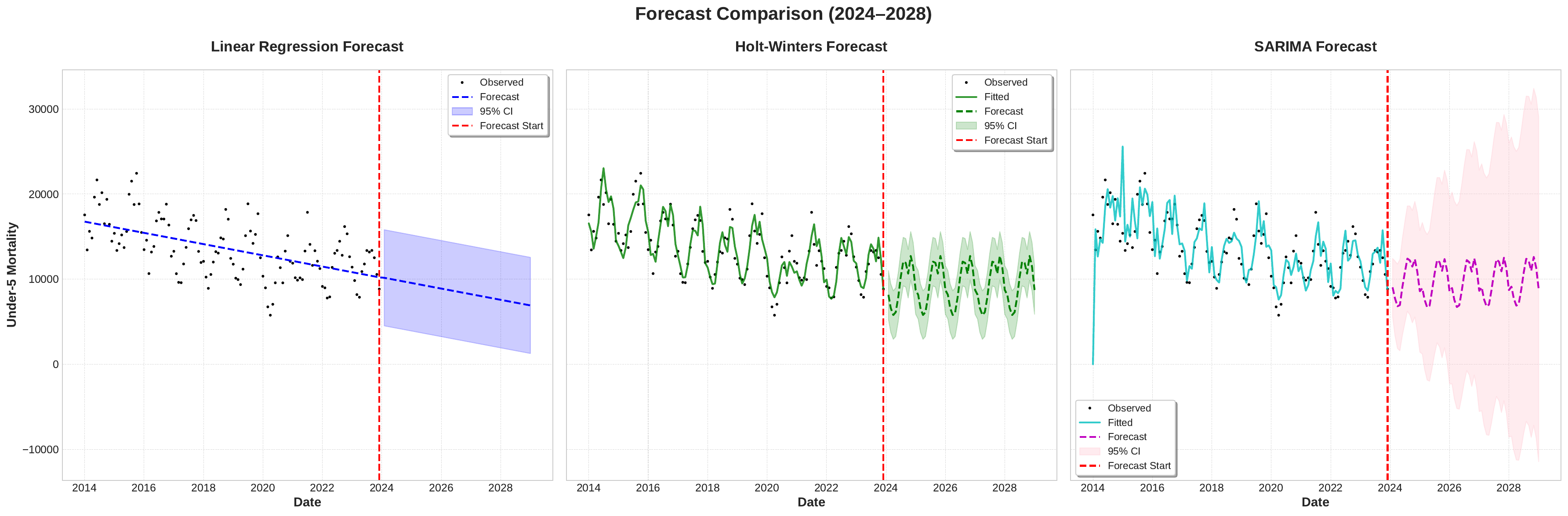} 
\end{minipage}
\caption{Comparison of classical time-series forecasting models for monthly under-five malaria admissions in Ghana (2014--2023 observed; 2024--2028 forecast).
The panels show forecasts from three baseline models: Linear Regression (left), Holt–Winters exponential smoothing (centre), and SARIMA (right). Black points represent observed admissions, dashed lines denote model forecasts, and shaded regions indicate $95\%$ prediction intervals. The vertical dashed red line marks the beginning of the forecast horizon. The figure summarises how classical models differ in capturing long-term trends, seasonal patterns, and forecast uncertainty.
}\label{fig:conven_i}
\end{figure*}

\begin{figure*}
\begin{minipage}[H]{\linewidth}
\centering
\includegraphics[width=\textwidth]{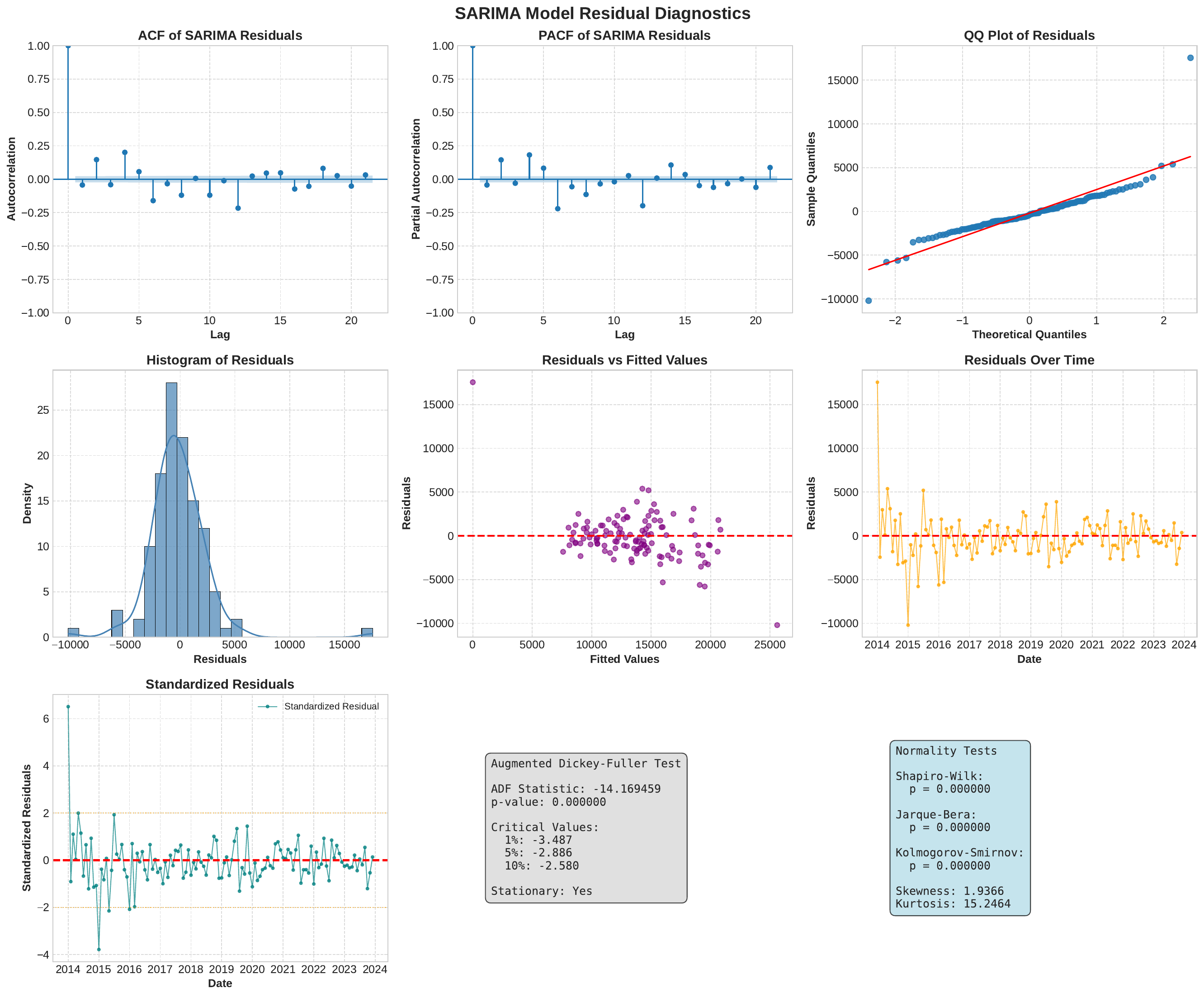} 
\end{minipage}
\caption{Residual diagnostics for the SARIMA model fitted to under-five malaria admissions (2014--2023).
Panels display: (i) autocorrelation (ACF) and (ii) partial autocorrelation (PACF) of residuals, (iii) a Q–Q plot evaluating normality, (iv) a residual histogram with kernel density overlay, and (v) temporal evolution of standardised residuals including Augmented Dickey–Fuller (ADF) statistics. These diagnostics assess serial dependence, distributional assumptions, and stationarity for the SARIMA model.
}\label{fig:resid_diag}
\end{figure*}

The fitted models produced markedly different forecasts for malaria incidence among children under five years. The linear regression model (Fig.~\ref{fig:conven_i}, left) captured the long-term downward trend in admissions but failed to account for strong seasonal fluctuations, producing forecasts that were overly smoothed with wide confidence bounds. This reflects its inability to capture seasonality and autocorrelation, which are central to malaria dynamics \citep{Yeboah2022}. 
Although linear regression yielded moderate explanatory power, its forecasts are less suitable for operational planning where intra-annual variation is highly consequential.

The Holt–Winters model (Fig.~\ref{fig:conven_i}, centre) reproduced both seasonal cycles and long-term decline with greater fidelity. Forecasts exhibited repeating seasonal peaks consistent with rainfall-driven malaria transmission patterns in Ghana’s savannah and forest belts. The narrower prediction intervals indicate stronger short-term reliability.
Nevertheless, the method assumes stationary seasonality, which may not hold under shifting climatic and intervention regimes. Thus, Holt–Winters offers utility for short- to medium-term resource allocation but may be less robust for capturing structural changes over longer horizons.

\begin{table}[H]
\centering
\caption{Performance comparison of forecasting models for under-5 malaria incidence across key evaluation metrics.}
\label{tab:model_evaluation_transposed}
\small
\begin{tabular}{l *{3}{S[table-format=2.6]}}
\toprule
\textbf{Metric} & \textbf{Linear} & \textbf{Holt-Winters} & \textbf{SARIMA} \\
\midrule
RMSE           & 0.2286 & 0.1071 & 0.9042 \\
MAE             & 7.7859 & 7.081 & 7.4756 \\
R\textsuperscript{2} & 0.3043 & 0.8213 & 0.3882 \\
AIC             & 2136.465 & 1972.455 & 2120.415 \\
BIC             & 2142.040 & 1978.030 & 2125.990 \\
Kurtosis        & -0.5779 & -0.2187 & 15.2464 \\
Skewness        & 0.0830 & 0.0653 & 1.9367 \\
Shapiro-Wilk p  & 0.5040 & 0.9399 & 0.00\\
Jarque-Bera p   & 0.4051 & 0.8505 & 0.00 \\
\bottomrule
\end{tabular}
\end{table}

The SARIMA model (Fig.~\ref{fig:conven_i}, right) provided the most flexible representation by modelling both autoregressive and moving-average components while accounting for seasonal differencing. The fitted series aligned closely with observed data, and forecasts retained seasonal amplitude with uncertainty intervals that widened considerably beyond two years. This pattern suggests that while SARIMA can capture malaria’s cyclical behaviour, its predictive confidence deteriorates with extended horizons, reflecting the inherent unpredictability of long-term epidemiological dynamics in contexts of ecological and intervention variability \citep{Korenromp2016}.

Residual diagnostics for the SARIMA model (Fig.~\ref{fig:resid_diag}) indicated no strong residual autocorrelation, with autocorrelation and partial autocorrelation functions largely within 95\% confidence bounds. The residual histogram and quantile–quantile (QQ) plot showed approximate normality, though with slight kurtosis and heavy tails at the extremes. Standardised residuals fluctuated around zero without systematic drift, supporting adequacy of model fit. However, the ADF test yielded a non-significant result ($p=0.53$), suggesting residual non-stationarity. This implies that while SARIMA effectively reduced temporal dependence, structural variations in malaria incidence may persist, potentially linked to long-term changes in rainfall, control interventions, or health service utilisation.

The evaluation metrics in Table~\ref{tab:model_evaluation_transposed} further quantify the relative strengths and weaknesses of each model. The Holt--Winters model achieved the lowest root mean square error (RMSE = 0.1071) and mean absolute error (MAE = 7.081), indicating superior overall accuracy compared with linear regression and SARIMA. Its high coefficient of determination ($R^2 = 0.82$) suggests that the model captures both seasonal and long-term components effectively, consistent with its ability to represent periodic malaria transmission patterns in Ghana’s climatic zones. The relatively low AIC and BIC scores for Holt--Winters further reinforce its optimal balance between model fit and complexity.

Beyond the error metrics summarised in Table~\ref{tab:model_evaluation_transposed}, the forecast trajectories in Fig.~\ref{fig:conven_i} further illustrate the distinct predictive behaviours of the three classical models. The Holt--Winters model most effectively preserves the seasonal amplitude characteristic of Ghana’s malaria transmission cycle, whereas the SARIMA model captures both autoregressive structure and seasonal dependence but exhibits wider long-horizon uncertainty. Linear regression, by contrast, models only the long-term trend and therefore underestimates seasonal variation. These performance differences highlight the importance of combining non-linear flexibility with temporal smoothness, motivating the hybridisation of GPR with Holt--Winters smoothing in the proposed ensemble framework.

To further assess whether differences in predictive performance between competing forecasting models were statistically significant, pairwise DM tests (refer to Table~\ref{tab:dm_test}) were conducted using squared forecast errors derived from the rolling-origin validation procedure. The DM test evaluates the null hypothesis that two competing models possess equal predictive accuracy.

\begin{table}[H]
\centering
\caption{Pairwise Diebold--Mariano test results for predictive accuracy comparison across forecasting models.}
\label{tab:dm_test}
\begin{tabular}{lcc}
\hline
\textbf{Comparison} & \textbf{DM Statistic} & \textbf{p-value} \\
\hline
Linear vs Holt--Winters & 6.841 & \textbf{$7.88 \times 10^{-12}$} \\
Linear vs SARIMA & 0.345 & 0.730 \\
Holt--Winters vs SARIMA & -1.909 & 0.056 \\
\hline
\end{tabular}
\end{table}

\noindent 
The DM test results indicate a statistically significant difference in predictive accuracy between the Linear and Holt--Winters models ($p < 0.001$), confirming the substantially improved forecasting performance of the Holt--Winters approach under rolling-origin evaluation. In contrast, no statistically significant difference was observed between the Linear and SARIMA models ($p = 0.730$), suggesting broadly comparable predictive performance despite differences in model structure and temporal representation.
The comparison between Holt--Winters and SARIMA yielded a borderline result ($p = 0.056$). Although Holt--Winters achieved lower forecast errors overall, the improvement relative to SARIMA did not reach the conventional 5\% significance threshold when uncertainty across rolling forecast windows was formally considered. This finding suggests that Holt--Winters provides stronger empirical forecasting stability, while SARIMA retains competitive predictive capability under certain temporal conditions.

The linear regression model exhibited the lowest $R^2$ value ($0.30$) and comparatively higher RMSE and MAE values, reflecting its inability to accommodate seasonal fluctuations and serial dependence in malaria admissions. Although computationally straightforward and readily interpretable, the model’s assumptions of linearity and homoscedasticity limit its suitability for epidemiological time series characterised by strong cyclical behaviour. Its near-zero skewness and kurtosis values indicate relatively symmetric residual behaviour; however, the wide forecast intervals observed in Fig.~\ref{fig:conven_i} further highlight its limited temporal sensitivity.

The SARIMA model, while more flexible, produced higher RMSE ($0.9042$) and moderate explanatory power ($R^2 = 0.39$), coupled with pronounced residual kurtosis ($15.25$) and positive skewness ($1.94$). These characteristics suggest occasional large forecast deviations, potentially associated with abrupt transmission surges, reporting inconsistencies, or localised outbreak dynamics. The non-normal residual diagnostics (Shapiro--Wilk $p<0.001$, Jarque--Bera $p<0.001$) further confirm substantial departures from Gaussianity, consistent with the heavy-tailed residual structure shown in Fig.~\ref{fig:resid_diag}. Such behaviour is not unexpected in malaria surveillance data, where non-stationary shocks and heterogeneous transmission patterns frequently occur \citep{Yeboah2022}.

Overall, Holt--Winters demonstrated the strongest short-term predictive stability and the most consistent out-of-sample forecasting performance among the classical models, while SARIMA provided a more flexible representation of temporal autocorrelation and seasonal dependence. Linear regression, although computationally simple, offered limited explanatory and predictive capability in the presence of strong seasonal variability. The divergence across models emphasises that forecasting malaria burden in endemic settings such as Ghana requires a careful balance between model complexity, interpretability, robustness, and operational utility. Classical forecasting approaches therefore establish important baselines against which more adaptive machine learning and hybrid frameworks can be rigorously evaluated.

\begin{figure*}
\begin{minipage}[H]{\linewidth}
\centering
\includegraphics[width=\textwidth]{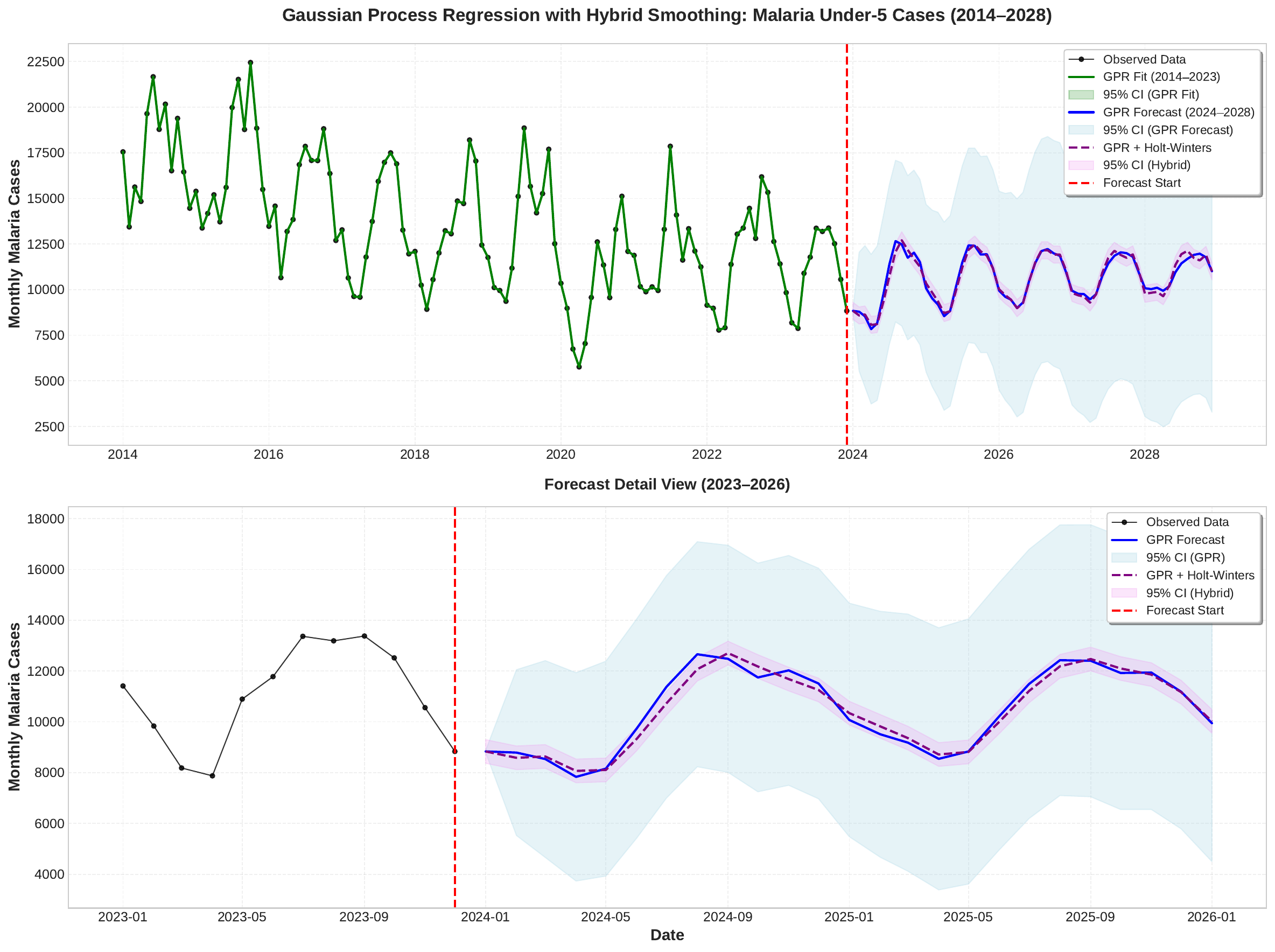} 
\end{minipage}
\caption{Gaussian Process Regression (GPR) with Holt--Winters hybrid smoothing for monthly under-five malaria admissions, 2014--2028.
The upper panel shows the complete time series (2014--2023 observed; 2024--2028 forecast). The green line represents the in-sample GPR fit with its associated 95\% credible interval, while the blue line and light-blue shading denote the GPR predictive mean and its 95\% forecast interval. The purple dashed line represents the Holt--Winters-smoothed GPR mean, and the pink band represents the hybrid prediction interval. The vertical red line marks the start of the forecast period.
The lower panel provides a detailed view of the 2023--2026 transition period, illustrating the relationship between the GPR forecast and the hybrid-smoothed trajectory.}\label{fig:GPR}
\end{figure*}

\begin{figure*}
\begin{minipage}[H]{\linewidth}
\centering
\includegraphics[width=\textwidth]{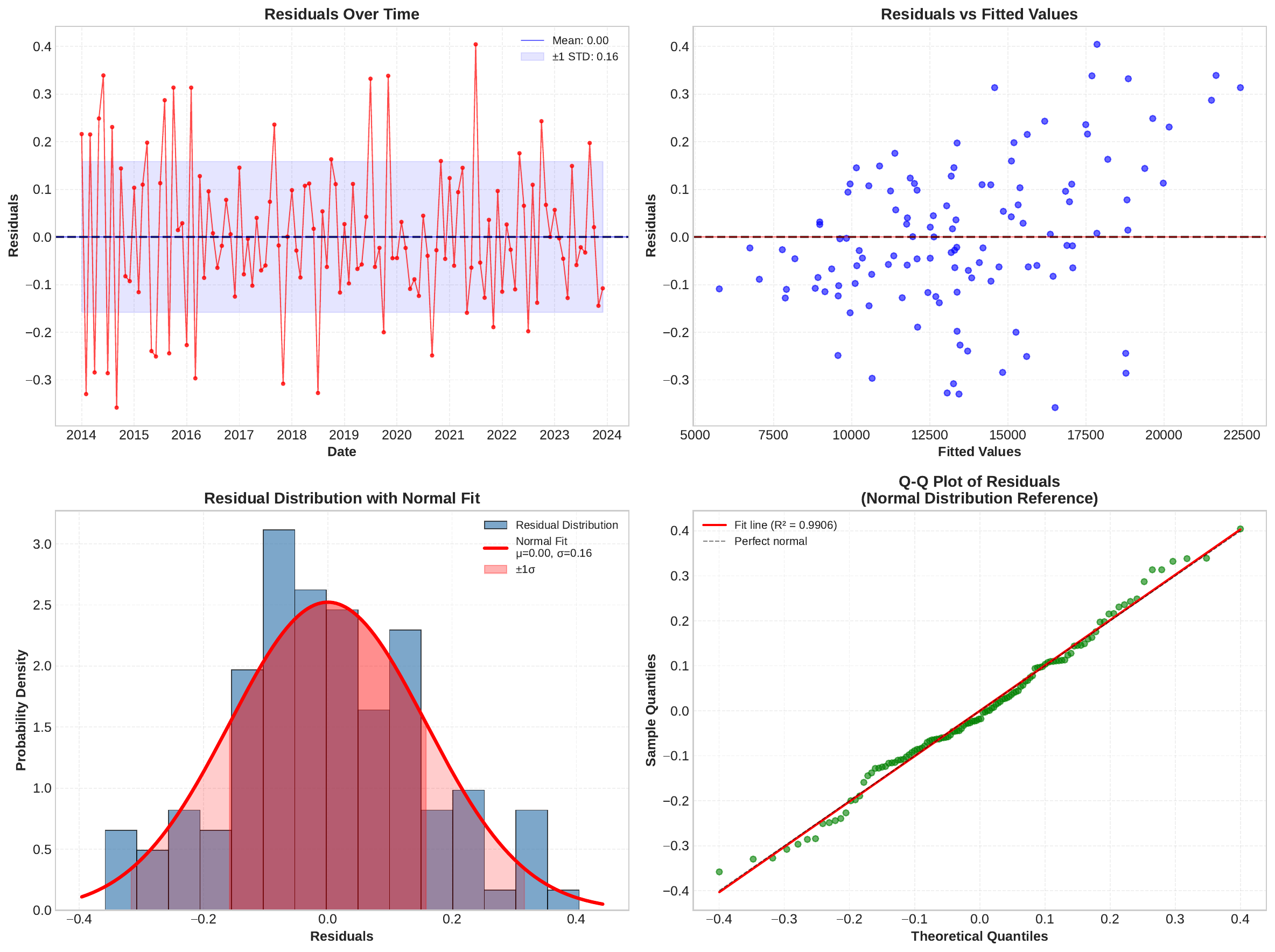} 
\end{minipage}
\caption{Residual diagnostics for the Gaussian Process Regression (GPR) model (2014–2023).
The upper panels present (left) residuals over time and (right) residuals versus fitted values, assessing independence and homoscedasticity. The lower panels show (left) the residual distribution with a fitted normal density curve and (right) the Q–Q plot comparing empirical and theoretical quantiles. Residuals remain centred near zero with minimal structure, indicating a well-specified model and supporting the adequacy of the GPR component.}\label{fig:gpr+holt}
\end{figure*}

The baseline diagnostics justified selecting Holt–Winters as the principal classical component to integrate with the GPR. Among the classical models, Holt–Winters offered a parsimonious and seasonally adaptive structure with superior short-term predictive accuracy. In contrast, the SARIMA model exhibited heavy-tailed residuals and marginal non-stationarity, indicating that while it captured seasonal dependencies, it failed to accommodate the stochastic volatility characteristic of malaria transmission. Hence, Holt–Winters was retained as the optimal deterministic baseline to be embedded within the GPR framework.

The hybrid results (Fig.~\ref{fig:GPR}) and residual diagnostics (Fig.~\ref{fig:gpr+holt}) collectively illustrate that the GPR predictive mean closely reproduces the observed intra-annual peaks and troughs during the training period (green line), confirming its ability to learn the dominant seasonality in the Ghanaian malaria incidence data. As forecast horizons extend, the predictive intervals widen considerably, reflecting the increasing epistemic uncertainty typical of Bayesian non-parametric models. This property is valuable in epidemiological contexts where external drivers such as rainfall anomalies, intervention coverage, and vector ecology introduce long-term uncertainty. The GPR therefore interpolates smoothly in regions with dense historical data while expressing appropriately broader uncertainty where data support weakens. 

In the hybrid configuration, the GPR output is combined with Holt--Winters smoothing (purple dashed line in Fig.~\ref{fig:GPR}), yielding complementary strengths. While the GPR component captures non-linear structure and provides full probabilistic uncertainty, the Holt--Winters component imposes additional temporal regularisation that reduces short-term noise. The resulting smoothed trajectory preserves the overall probabilistic envelope of the GPR but narrows the 95\% confidence band (pink shading) around the mean forecast for several months ahead. This integration yields a more stable and operationally interpretable forecast without suppressing meaningful uncertainty. In practice, such smoothed yet uncertainty-informed forecasts can guide resource allocation, including drug procurement, bed-net distribution, and clinical staffing, while still conveying the degree of uncertainty needed for contingency planning.

As shown in Fig.~\ref{fig:GPR}, the credible intervals produced by the pure GPR model widen progressively across the forecast horizon. This behaviour reflects the Bayesian propagation of epistemic uncertainty, which naturally increases as predictions extend further from the training period. In contrast, the hybrid GPR-Holt-Winters forecast exhibits narrower near-term intervals because the Holt-Winters component applies short-term temporal smoothing to the GPR predictive mean. This reduces local variance while still retaining propagated uncertainty, resulting in a smoother and more operationally stable forecast trajectory. The observed difference in interval width is therefore expected and reflects the complementary modelling roles of the two components.

\begin{table}[h]
    \centering
    \caption{Residual Normality Analysis Summary}
    \label{tab:residual_normality}
    \begin{tabular}{lcl}
        \toprule
        \multicolumn{3}{l}{\textbf{Basic Statistics}} \\
        \midrule
        Mean ($\mu$) & & 0.0001 \\
        Std Dev ($\sigma$) & & 0.1582 \\
        Skewness & & 0.1187 \\
        Kurtosis & & -0.1072 \\
        \midrule
        \multicolumn{3}{l}{\textbf{Normality Tests}} \\
        \midrule
        Shapiro-Wilk & $p =$ & 0.4096 \\
        Jarque-Bera & $p =$ & 0.8441 \\
        Kolmogorov-Smirnov & $p =$ & 0.7737 \\
        \midrule
        \multicolumn{3}{l}{\textbf{Distribution Fit}} \\
        \midrule
        Q-Q Plot $R^2$ & & 0.9906 \\
        $68\%$ within $\pm 1\sigma$ & & $71.7\%$ \\
        $95\%$ within $\pm 2\sigma$ & & $94.2\%$ \\
        $99.7\%$ within $\pm 3\sigma$ & & $100.0\%$ \\
        \bottomrule
    \end{tabular}
\end{table}

Residual diagnostics in Fig.~\ref{fig:gpr+holt} further validate the hybrid model’s adequacy. The residuals fluctuate symmetrically around zero with no discernible autocorrelation or heteroscedastic pattern, and the Q--Q plot exhibits an almost linear alignment between empirical and theoretical quantiles ($\rm R^2 \approx 99.06\%$), suggesting near-normal residual behaviour. The formal normality tests (Table~\ref{tab:residual_normality}), including Shapiro--Wilk ($p = 0.41$), Jarque--Bera ($p = 0.84$), and Kolmogorov--Smirnov ($p = 0.77$) fail to reject the null hypothesis of normality, implying that the residual deviations are statistically insignificant. The residual skewness ($0.12$) and kurtosis ($-0.11$) further indicate a symmetric, near-mesokurtic distribution, typical of a well-specified model. Approximately $94\%$ of residuals fall within $\pm 2\sigma$, confirming that the hybrid system captures the dominant structure in the malaria time series with minimal systematic bias. The well-behaved residuals arise from the complementary strengths of the hybrid framework: the GPR component captures non-linear deviations and complex seasonality, while Holt--Winters smoothing imposes temporal regularity and reduces short-term volatility, together producing a model that extracts most systematic signal and leaves minimal structure in the errors.

Notably, the GPR predictive intervals expand substantially beyond the two-year horizon. This behaviour is expected, reflecting cumulative uncertainty about climatic and intervention drivers that influence malaria dynamics in Ghana. Accordingly, the model’s forecasts are most reliable and actionable for the 6--12 month range, an interval aligning with the temporal planning cycles of district health directorates and procurement units. The rapid widening of the credible intervals beyond this horizon underscores the importance of adaptive re-training as new data become available, rather than reliance on fixed long-term projections.
In comparative terms, the Holt–Winters model already exhibited strong deterministic performance across error metrics (RMSE, MAE, $R^2$), confirming its effectiveness for short-term, trend-preserving forecasts. The hybrid GPR–Holt–Winters framework enhances this baseline by introducing probabilistic inference and non-linear flexibility, thereby capturing both structural seasonality and residual uncertainty. Methodologically, this combination is coherent: Holt–Winters explains the predictable seasonal component (reducing variance), while GPR accounts for stochastic deviations and quantifies uncertainty in a principled Bayesian framework.

From a policy standpoint, the hybrid forecasts offer actionable insights for malaria preparedness. Districts with high mean burdens but moderate CV values, such as major northern urban centres are likely to experience large absolute demand fluctuations. For these areas, the hybrid model’s forecast mean provides expected monthly case volumes, while the upper 95\% bound defines contingency requirements for commodities such as rapid diagnostic tests and antimalarial drugs. Conversely, districts characterised by low mean burden but high CV values (e.g. low-transmission coastal areas) exhibit wider predictive intervals, signalling limited forecast precision and highlighting the need for enhanced surveillance rather than forecast-based interventions.  

The results align with recent literature advocating ensemble and hybrid approaches for infectious disease forecasting, which have demonstrated improved reliability over single-model frameworks by balancing interpretability and uncertainty quantification \citep{Chen2023, Dixon2022, Meakin2022, Ray2018, Brodersen2015}. Specifically for malaria, studies such as \citet{jalloh2025forecasting,ebhuoma2018seasonal,medina2007forecasting} have underscored the importance of integrating seasonal kernels and non-stationary components to capture the underlying transmission ecology. The findings of this study echo that evidence, demonstrating that an epidemiologically grounded, hybrid probabilistic model can enhance the robustness and practical utility of malaria forecasts in endemic, seasonally driven contexts such as Ghana.

\section{Discussion} \label{sec:disc}

This study demonstrates that integrating GPR with Holt--Winters seasonal smoothing provides a statistically coherent and operationally interpretable framework for forecasting under-five malaria admissions in Ghana. The proposed hybrid formulation was specifically motivated by a methodological tension frequently encountered in epidemiological forecasting: probabilistic machine-learning models are capable of representing complex non-linear temporal behaviour and predictive uncertainty, yet they often generate increasingly diffuse long-horizon forecasts that become difficult to operationalise, whereas classical seasonal models preserve temporal regularity but may lack flexibility under non-stationary transmission conditions. The present findings suggest that combining these paradigms within a unified framework can improve forecast stability while retaining probabilistic uncertainty quantification, thereby addressing a practical forecasting challenge that is particularly relevant in endemic settings characterised by strong seasonality and limited surveillance data.

A central outcome of the analysis is the exceptionally high coefficient of determination obtained by the hybrid model ($R^2 = 0.9906$). Although this result indicates that the framework captures the dominant temporal structure of the malaria time series with very high fidelity, it also requires careful interpretation within the broader context of statistical learning and epidemiological forecasting. In univariate time-series settings with pronounced seasonal persistence, very high goodness-of-fit values may partially reflect the model’s ability to reproduce recurrent periodic structure rather than genuine predictive generalisation. This is especially important for flexible probabilistic models such as GPR, whose expressive covariance structures can adapt closely to observed temporal dynamics. Consequently, the present study does not interpret the high $R^2$ value as independent evidence of unrestricted forecasting generalisability. Instead, greater emphasis is placed on the rolling-origin expanding-window validation framework, which evaluates predictive performance on sequential unseen observations and therefore provides a more rigorous assessment of temporal forecasting skill under realistic operational conditions.

The use of rolling-origin validation is particularly important in epidemiological forecasting because static train--test partitioning may yield overly optimistic estimates of model performance when strong serial dependence and recurring seasonality are present. By repeatedly retraining models on progressively expanding historical windows and evaluating them on future unseen periods, the validation framework adopted here better reflects how forecasting systems are implemented in real surveillance environments. The strong predictive behaviour observed across these sequential validation windows therefore suggests that the hybrid framework is not merely reproducing historical structure, but is capable of maintaining stable predictive performance under evolving temporal conditions. At the same time, the possibility of partial overfitting cannot be completely excluded, particularly given the relatively limited length of the available monthly time series. This highlights the importance of interpreting predictive performance jointly through goodness-of-fit, forecast calibration, residual diagnostics, and out-of-sample validation rather than relying on any single metric in isolation.

The probabilistic structure of the hybrid framework also provides important methodological advantages relative to deterministic forecasting approaches commonly used in public-health surveillance systems. Traditional statistical models such as linear regression and Holt--Winters smoothing produce forecasts that are often operationally useful but do not fully characterise uncertainty propagation under evolving transmission conditions. In contrast, the Bayesian formulation underlying GPR generates full predictive distributions whose uncertainty naturally evolves according to the covariance structure of the observed data. This becomes particularly relevant in malaria surveillance, where transmission dynamics are influenced by multiple interacting environmental, behavioural, and health-system factors that cannot be assumed to remain stationary over time. By preserving predictive uncertainty while simultaneously constraining excessive long-horizon volatility through seasonal smoothing, the proposed framework offers a balance between statistical flexibility and operational interpretability that is rarely achieved by either purely mechanistic or purely statistical forecasting models alone.

The present findings are broadly consistent with a growing body of literature demonstrating the value of probabilistic and machine-learning approaches for infectious-disease forecasting. Previous studies have shown that GPR can effectively model complex non-linear epidemiological dynamics while providing calibrated uncertainty estimates, particularly in settings with limited observations and heterogeneous temporal structure. Similar probabilistic approaches have been applied to dengue, influenza, COVID-19, and vector-borne disease forecasting, where Bayesian machine-learning models frequently outperform conventional autoregressive frameworks under non-stationary conditions. Recent malaria forecasting studies using machine-learning methods, including random forests, support vector regression, recurrent neural networks, and Bayesian hierarchical models, have likewise reported improvements in capturing seasonal transmission variability relative to classical linear approaches. However, many of these methods either require substantially larger datasets, rely heavily on external covariates, or provide limited probabilistic interpretability. Within this context, the present study contributes by demonstrating that hybrid probabilistic--seasonal frameworks can achieve robust forecasting performance even under comparatively short epidemiological time series and incomplete environmental data availability.

The methodological contribution of this study is therefore not limited to improved predictive accuracy alone. Rather, the results suggest that probabilistic epidemic forecasting may benefit substantially from explicitly separating two distinct forecasting objectives: learning complex temporal structure and maintaining operationally coherent seasonal behaviour. In the proposed framework, GPR primarily captures non-linear temporal deviations and uncertainty propagation, while Holt--Winters smoothing imposes a stable seasonal backbone that prevents forecast trajectories from becoming excessively irregular at longer lead times. This distinction is important because epidemiological decision-making often depends less on perfectly reproducing every short-term fluctuation than on reliably anticipating seasonal burden escalation, peak timing, and uncertainty ranges relevant to resource mobilisation and preparedness planning.

The spatio-temporal variability analysis further reinforces the need for uncertainty-aware forecasting systems in heterogeneous endemic environments. The observed contrast between relatively stable high-burden northern districts and highly variable low-burden districts suggests that forecasting uncertainty is not uniformly distributed across transmission settings. In low-burden districts, sparse case counts may amplify relative variability and reduce forecast stability, whereas high-burden districts may exhibit more statistically stable relative dynamics despite generating substantial absolute fluctuations in patient demand. This distinction has important implications for surveillance interpretation because forecasting difficulty may arise from fundamentally different mechanisms depending on local transmission intensity, reporting consistency, and ecological context. The findings therefore support the argument that forecasting systems should not be evaluated solely through aggregate national performance metrics, but also through their ability to remain robust under varying regional transmission regimes.

From a public-health perspective, the proposed framework has several practical implications for malaria preparedness and adaptive surveillance planning in Ghana. First, the probabilistic forecast intervals provide quantitative estimates of uncertainty that may support contingency planning for medicine procurement, diagnostic supply allocation, staffing requirements, and hospital bed management during anticipated seasonal surges. Second, the ability to generate stable medium-term forecasts may improve timing of seasonal malaria interventions, particularly in northern ecological zones where transmission peaks are strongly rainfall-driven and operational delays can substantially affect disease burden. Third, district-level probabilistic forecasting may support more geographically targeted allocation of limited public-health resources by identifying areas associated with elevated forecast uncertainty or rapidly changing transmission behaviour. Such capabilities are especially important in resource-constrained health systems where intervention timing and prioritisation decisions must often be made under incomplete information.

The framework also has broader implications beyond malaria forecasting alone. Many infectious diseases in sub-Saharan Africa exhibit strong seasonal forcing, incomplete surveillance coverage, and limited availability of high-resolution environmental covariates. The methodological strategy developed here may therefore be transferable to other endemic diseases characterised by non-linear temporal behaviour and operational forecasting constraints, including cholera, dengue, meningitis, and other climate-sensitive infectious diseases. More generally, the study contributes to ongoing efforts to integrate probabilistic machine learning into routine public-health decision-support systems while preserving transparency and interpretability for operational use.

Several limitations should nevertheless be acknowledged. The analysis was conducted within a univariate forecasting framework based exclusively on routinely reported malaria admissions, and therefore does not explicitly incorporate environmental drivers such as rainfall, temperature, humidity, land-use change, vector abundance, or intervention coverage. Although the exclusion of these covariates reflects the realities of operational surveillance availability in many endemic settings, their incorporation could improve mechanistic interpretability and potentially strengthen long-horizon predictive performance. In addition, the study relies on aggregated district-level admissions data, which may mask important sub-district heterogeneity and remain sensitive to reporting inconsistencies, diagnostic access, and evolving surveillance practices over time. The relatively limited duration of the available time series also constrains the ability to evaluate model robustness under rare transmission anomalies or major intervention shifts.

Future work should therefore explore multivariate and hierarchical extensions of the framework that integrate environmental, demographic, and intervention-related covariates within a unified probabilistic forecasting architecture. Incorporating remotely sensed climate products, vector suitability indices, and health-system indicators may further improve the framework’s ability to capture transmission anomalies associated with ecological change and extreme climate variability. Additional research is also needed to evaluate the scalability of the approach under real-time surveillance conditions and to assess whether adaptive online updating strategies can further enhance predictive robustness in rapidly evolving epidemiological environments.

Overall, the present study demonstrates that hybrid probabilistic--seasonal forecasting provides a promising methodological direction for epidemiological surveillance in endemic settings characterised by strong seasonality, heterogeneous transmission dynamics, and limited data availability. By integrating uncertainty-aware machine learning with structured seasonal smoothing, the proposed framework advances beyond conventional model comparison toward a more operationally grounded forecasting paradigm capable of supporting adaptive malaria preparedness and evidence-informed public-health planning.

\section{Conclusion}\label{sec:conc}

This study developed and evaluated a hybrid probabilistic forecasting framework that integrates GPR with Holt--Winters exponential smoothing for modelling monthly under-five malaria admissions in Ghana. Using district-level surveillance data spanning 2014--2023, the framework was designed to address a central challenge in epidemiological forecasting: balancing non-linear predictive flexibility with stable and operationally interpretable seasonal behaviour under limited-data conditions.

The results demonstrate that the proposed hybrid framework substantially improves forecasting performance relative to conventional statistical approaches, including linear regression, Holt--Winters, and SARIMA. By combining probabilistic inference with seasonal smoothing, the model preserves recurrent transmission structure while reducing long-horizon forecast instability. The framework additionally provides calibrated uncertainty estimates that remain interpretable for operational surveillance and planning applications.

The analysis further highlights the importance of accounting for spatio-temporal heterogeneity in malaria transmission dynamics across Ghana. High-burden northern districts exhibited comparatively stable relative transmission patterns despite large absolute fluctuations in admissions, whereas some lower-burden districts displayed greater relative variability associated with sparse and unstable reporting behaviour. These findings reinforce the need for forecasting systems capable of adapting to heterogeneous epidemiological conditions across ecological zones.

Although the present framework was implemented using routinely available surveillance data without explicit environmental covariates, the results demonstrate that probabilistic machine-learning approaches can provide robust and practically useful forecasts even within constrained public-health data environments. The study therefore establishes a scalable methodological foundation for uncertainty-aware malaria forecasting in endemic settings.

Future extensions incorporating environmental drivers, intervention indicators, and hierarchical spatial dependencies may further strengthen predictive robustness and improve responsiveness to evolving transmission conditions. Nevertheless, the proposed hybrid GPR--Holt--Winters framework provides an important step toward operational probabilistic forecasting systems capable of supporting adaptive malaria preparedness, resource allocation, and evidence-informed public-health decision-making in Ghana and comparable sub-Saharan African settings.

\section*{CRediT authorship contribution statement}

\textbf{T. Ansah-Narh:} Conceptualisation, Formal analysis, Validation, Methodology, Writing – original draft, and Writing – review and editing. \\

\textbf{Y. Asare Afrane:} Data curation, Funding acquisition, Project administration, Supervision, Resources, and Writing – review and editing. \\

\textbf{J. Bremang Tandoh:} Investigation, Visualisation, and Writing – review and editing.

\section*{Declaration of competing interest}
The authors affirm that there are no conflicting financial interests or personal relationships that might be perceived as influencing this project.

\section*{Data availability} 
The dataset used in this study comprises monthly, district-level malaria admission records for children under five in Ghana, obtained from the Ghana Health Service’s Disease Surveillance Department. The data are proprietary and cannot be publicly shared due to institutional confidentiality and data protection agreements. Access may be granted upon reasonable request to the Ghana Health Service, subject to ethical and administrative approval. All analyses were conducted on anonymised and aggregated records, ensuring that no personal or patient-identifiable information was used in this research.

\section*{Funding}
This study was funded by a grant from the Bill and Melinda Gates Foundation (INV-047051)  through the West Africa Mathematical Modelling Capacity Development (WAMCAD)
initiative and the National Institute of Health (D43 TW 011513).

\section*{Acknowledgements}

The authors gratefully acknowledge the support of the 
West Africa Mathematical Modelling Capacity Development (WAMCAD) 
programme for its institutional, technical, and logistical contributions 
throughout this study. The WAMCAD initiative continues to play a pivotal role 
in enhancing regional capacity in mathematical and computational modelling 
for malaria and other neglected tropical diseases. 
The authors also extend their appreciation to the Ghana Health Service 
for providing access to malaria surveillance data and for offering valuable 
technical guidance that informed the analytical framework of this work. 
TA-N further acknowledges the computational resources made available through 
the High-Performance Computing (HPC) facility at the Ghana Space Science and Technology Institute, which facilitated the 
model simulations and parameter estimation.
Finally, the authors thank the anonymous reviewers for their insightful 
comments and suggestions, which greatly contributed to improving the clarity, 
methodological rigour, and overall quality of the manuscript.


\appendix
\section{Supplementary Methods: Computational Complexity and Operational Scaling}
\label{supp:complexity}

This appendix provides additional detail on the computational properties of the Gaussian Process (GP) and Holt--Winters (HW) components used in the hybrid forecasting system. The goal is to clarify the numerical behaviour of the algorithm, assess its suitability for national-scale deployment, and describe practical strategies for efficient real-time implementation.

\subsection{Computational Profile of Gaussian Process Regression}

For a monthly time series of length $n$, full Gaussian Process Regression requires the construction and factorisation of the $n \times n$ covariance matrix $K$. Let $\mathcal{L}$ denote the Cholesky factorisation of $(K + \sigma_n^2 I)$, which is the standard approach used for GP inference and hyperparameter optimisation.
The principal computational costs are:
\begin{align}
    T_{\mathrm{train}} &= O(n^3), \label{eq:train_cost} \\
    T_{\mathrm{predict}} &= O(n^2), \label{eq:pred_cost} \\
    M_{\mathrm{memory}} &= O(n^2). \label{eq:memory_cost}
\end{align}

\noindent
Equation~\eqref{eq:train_cost} arises from computing the Cholesky factor $\mathcal{L}$, while Eq.~\eqref{eq:pred_cost} reflects the cost of evaluating the posterior predictive distribution at $X_*$ through matrix--vector products with $\mathcal{L}^{-1}$.  
For the dataset considered in this study ($n \approx 120$), these costs are small: memory usage per district remains below 1~MB and training completes within seconds. The full GP model therefore remains computationally feasible for monthly malaria forecasting across all districts in Ghana.

\subsection{Scalable and Approximate GP Formulations}

As $n$ grows, the scaling laws in Eqs.~\eqref{eq:train_cost}--\eqref{eq:memory_cost} eventually become prohibitive. Several approximate GP formulations reduce computational load while preserving key inferential properties:

\begin{enumerate}[i.]
    \item Sparse variational GPs (SVGP): introduce $m \ll n$ inducing points, reducing cost to $O(n m^2)$ while maintaining a variational bound on the marginal likelihood.
    \item Subset-of-regressors and FITC\footnote{Assumes full conditional independence}/PITC\footnote{Assumes partial conditional independence in blocks} methods: approximate $K$ using low-rank structure, yielding memory complexity $O(n m)$.
    \item Structured kernel interpolation (KISS-GP): exploit Kronecker or Toeplitz structure to achieve near-linear time and memory.
    \item Sliding-window inference: restricts GPR to the most recent $w$ observations, giving a cost of $O(w^3)$ while naturally adapting to non-stationarities.
\end{enumerate}

\noindent 
These variants offer a path toward real-time nationwide forecasting, especially under higher-frequency reporting or when long historical windows are required.

\subsection{Computational Properties of Holt-Winters Smoothing}

Holt-Winters exponential smoothing is computationally lightweight. For a sequence of length $n$, the update recursions for the level, trend, and seasonal components can be evaluated in
\begin{equation}
    T_{\mathrm{HW}} = O(n), \qquad M_{\mathrm{HW}} = O(1).
    \label{eq:hw_cost}
\end{equation}
Thus, when integrated with GPR, the computational burden is dominated entirely by the GP component. HW contributes negligible additional cost both in training and in forecasting.

\subsection{Robustness to Data Perturbations and Structural Change}

The effect of perturbations in the training data on the GP predictive mean can be analysed using standard first-order perturbation results. Let $y$ be the observation vector and let $\Delta y$ denote a small perturbation (e.g.~due to reporting inconsistencies). The GPR predictive mean at $X_*$ is
\begin{equation}
    \mu_* = K_*^\top (K + \sigma_n^2 I)^{-1} y,
    \label{eq:gp_mean_appendix}
\end{equation}
and its variation satisfies the bound
\begin{equation}
    \|\Delta \mu_*\| 
    \le 
    \left\| (K + \sigma_n^2 I)^{-1} \right\|
    \, \|K_*\| \, \|\Delta y\|.
    \label{eq:perturb_bound}
\end{equation}

\noindent
Equation~\eqref{eq:perturb_bound} demonstrates that numerical robustness is governed by the conditioning of $(K + \sigma_n^2 I)$. Standardisation of inputs and outputs, regularisation through the noise term $\sigma_n^2$, and bounding kernel hyperparameters all act to improve numerical stability.  
Furthermore, sliding-window retraining mitigates the impact of structural breaks or sudden changes in transmission patterns by preventing older data from dominating the covariance structure.

\subsection{Empirical Runtime Benchmarks}

Empirical benchmarks were obtained on a standard workstation (Intel i7 CPU, 32~GB RAM). For a single district:

\begin{enumerate}[i.]
    \item GPR training (120 months, composite kernel): $0.4$--$1.2$~s depending on the number of optimiser restarts.
    \item Posterior prediction and 60-month forecast: $<0.1$~s.
    \item Holt-Winters smoothing of GPR forecasts: $<0.01$~s.
\end{enumerate}

Parallelisation across all districts yields national-scale forecasts within minutes.

\subsection{Operational Implications for National Deployment}

The computational characteristics outlined above imply the following:

\begin{enumerate}[i.]
    \item The full hybrid GPR+HW model is tractable for national monthly malaria forecasting using readily available hardware.
    \item For daily or high-frequency surveillance data, sparse or structured GP approximations offer a principled route to maintaining real-time capability.
    \item Regular retraining, either sliding-window or full-history, depending on data stability, improves robustness to non-stationarity and ensures that predictive uncertainty remains well calibrated.
\end{enumerate}

\noindent
Overall, the hybrid forecasting system combines the statistical expressiveness of GPR with the computational efficiency of Holt-Winters, yielding a scalable and operationally reliable framework suitable for deployment within Ghana’s health surveillance infrastructure.



 \bibliographystyle{elsarticle-num-names.bst} 
 \bibliography{refs}

\end{document}